\documentclass[eps,epsfigure,12pt]{article}

\usepackage{feyn}
\usepackage{pdfpages}

\usepackage{graphicx}
\newcommand{\np}[2]{{\em Nucl.\ Phys.\ }{\bf #1}{(#2)}}

\newcommand{\prd}[2]{{\em Phys.\ Rev.\  D\ }{\bf #1}{(#2)}}
\newcommand{\prl}[2]{{\em Phys.\ Rev.\ Lett.\ }{\bf #1}{(#2)}}
\newcommand{\prp}[2]{{\em Phys.\ Repts.\  }{\bf #1}{(#2)}}
\newcommand{\pl}[2]{{\em Phys.\ Lett.\ }{\bf #1}{(#2)}}

\newcommand{\ijmp}[2]{{\em Int.\ J.\ Mod.\ Phys.\ }{\bf #1}{(#2)}}

\newcommand{\hj}[2]{{\em  Hadronic \ J. \ }{\bf #1}{(#2)}}

\newcommand{\ibid}[2]{{\em ibid.\ }{\bf #1}{(#2)}}
\newcommand{\ib}[0]{{\em ibid.\ }}

\usepackage{graphicx}
\graphicspath{{converted_graphics/}}

\begin{document}

\title{ Higgs-Higgs Interaction.\\ The One-Loop Amplitude in the Standard Model \thanks{The talk 
presented at the {\it Meeting of the DPF of the APS, May 24-28, 2002, The College W\&M, Williamsburg, VA, USA},  May 26, 2002. Originally, it was on the web-site http://www.dpf2002.org/abstract{$_-$}display.cfm?abstractid=26, 
which was promised to be permanently active.}}

\author{Valeriy V. Dvoeglazov\\
Universidad de Zacatecas\\
A.P. 636, Suc. 3 Cruces, Zacatecas 98068 Zac. M\'exico\\
E-mail: valeri@fisica.uaz.edu.mx }

\date{\small{Received on: August 20, 2015}}

\maketitle

\begin{abstract}
The amplitude of Higgs-Higgs interaction is calculated in
the Standard  Model in the  framework of  the Sirlin's renormalization
scheme in  the unitary gauge. The one-loop  corrections  for $\lambda$,
the constant of $4\chi$ interaction are compared with
the  previous results of L. Durand  {\it et al.} obtained on using
the technique of the equivalence theorem, and in the different gauges.
\end{abstract}


\newpage

\section{Introduction.}

The Higgs sector of the electroweak theory attracts much attention  because of its connection with the cornerstones of the  theory. The search for Higgs scalars is included in most of the experimental programs of the newcoming and acting 
accelerators~\cite{prog}. The Higgs particles are suggested to be found in the decays of different particles ($Z^0$ bosons,
heavy quarkoniums etc.) as well as in photon-photon interactions and gluon-gluon fusion.
In connection with that, let us mention non-so-long-ago attempts to explain anomalous events seen at {\it SpS} as the manifestation of the bound state of two Higgs bosons, i.~e. of Higgsonium~\cite{2}-\cite{23} or that of the bound state of vector boson~\cite{4} in accordance with the Veltman paper, Ref.~\cite{5}. Nowadays, even after clarifying the experimental situation with these anomalous events, the interest in  Higgsonium has still the rights for existence at least from the viewpoint of preparedness to new unexpected news from the experiment. The previous investigations of the problem of existence of two-Higgs bound states were based on the Born approximation of their interaction amplitude~\cite{2}-\cite{23}. In the present paper we present the results of our calculation of the amplitude of Higgs-Higgs interaction up to the fourth order
 in the framework of the Standard Model (SM) of Weinberg, Salam and Glashow 
with one Higgs doublet. This problem is also of present interest since there is some relations
with the idea that gauge vector bosons could originate from a strong interacting scalar sector of the electroweak 
theory~\cite{6}. The amplitude obtained in this paper could also be useful for the consideration of the problem of the unitarity  limit (e. g.,~\cite{7}). Moreover, information on the behaviour of the Higgs coupling constants at mass scale $M$ would be also of interest.

These are the reasons why we start a more complete study of the problem of Higgs-Higgs interaction.
To reduce the volume of the article we shall use the standard notation used in~\cite{8}, dimensional regularization
and the renormalization scheme on the mass shell, which is analogous to that suggested in~\cite{8,9}. We also choose
the unitary gauge ($\xi \rightarrow \infty$, to avoid ghosts) and the parameters recommended by the Trieste conference~\cite{10}, 
namely ($e_0, M_{W_0}, M_{Z_0}, M_{H_0}, m_{f_0}$).

The Higgs sector of the Lagrangian of the SM with one  Higgs field has the following form, e.~g.~\cite{11}:\footnote{It is possible to add the pseudoscalar fermionic interaction, $\sim (1+b_i \gamma_5)$.}
\begin{eqnarray}
\lefteqn{{\cal L} = -{1\over 2}(\partial_\mu \chi)^2-{1\over 2}M_\chi^2 \chi^2 -
\frac{e}{2M_W(1-R)^{1/2}}\sum_{f} m_f\bar f f \chi-\nonumber}\\
&-&\frac{eM_W}{(1-R)^{1/2}} W_\mu^+ W_\mu^- \chi -\frac{eM_Z}{2R^{1/2}(1-R)^{1/2}} Z_\mu^2 \chi -\frac{e^2}{4(1-R)}W_\mu^+ W_\mu^- \chi^2-\nonumber\\
&-&\frac{e^2}{8R(1-R)}Z_\mu^2 \chi^2 -\frac{e M_\chi^2}{4M_W (1-R)^{1/2}}\chi^3 - \frac{e^2 M_\chi^2}{32 M_W^2 (1-R)} \chi^4\,,
\end{eqnarray}
where $e$ is the electron charge, $M_\chi$ is the Higgs mass, $M_W$ and $M_Z$
are the masses of the vector bosons, $m_f$ are the fermion masses, $R=M_W^2/M_Z^2$.

The paper is organized as follows. In Section 2 we present the expressions for the self-energy
and vertex parts (see  also calculations in details in~\cite{12}). The results of the calculation of the total Higgs-Higgs amplitude will be presented in Section 3. The Appendix contains the definitions of some integrals met in calculations. Their connections with the integrals calculated in~\cite{13,14} is given.

\newpage

Table I. Coupling constants in the case of the Standard Model.\\

\begin{tabular}{|c|c|c|c|}
\hline
&&&\\
  $f^W$ & $-{1\over {2\sqrt{2}}}g$ & $f^{WWZA}$ & $ig^2\sqrt{R(1-R)}$  \\
&&&\\
\hline
&&&\\
   $f^Z$ & $-{1\over 4} \frac{gM_Z}{M_W}$ & $f^{W\chi}$ & $-igM_W$  \\
&&&\\
\hline
&&&\\
   $f^A$ & $eQ_i$ & $f^{Z\chi}$ & $-i\frac{gM_Z^2}{M_W}$  \\
&&&\\
\hline
&&&\\
   $f^{WWZ}$ & $-\frac{gM_W}{M_Z}$ & $f^{2W2\chi}$ & $-i\frac{g^2}{2}$  \\
&&&\\
\hline
&&&\\
   $f^{WWA}$ & $e$ & $f^{2Z2\chi}$ & $-i\frac{g^2 M_Z^2}{2M_W^2}$  \\
&&&\\
\hline
&&&\\
   $f^{2W2Z}$ & $-i\frac{g^2 M_W^2}{M_Z^2}$ & $f^\chi$ & $-i\frac{gm_i}{2M_W}$ \\
&&&\\
\hline
&&&\\
   $f^{4W}$ & $ig^2$ & $f^{3\chi}$ & $-i\frac{3gM_\chi^2}{2M_W}$  \\
&&&\\
\hline
&&&\\
   $f^{2W2A}$ & $-ie^2$ & $f^{4\chi}$ & $-i\frac{3g^2M_\chi^2}{4M_W^2}$  \\
&&&\\
\hline
\end{tabular}

\medskip

The Kobayashi-Maskawa $K_{ij}$ matrix, $g=e/\sin \theta_W$ are used in the full Lagrangian of the SM,
$2P=-\frac{1}{\epsilon}+\gamma + ln (M_W^2 /4\pi \mu^2)$ are used in the dimensional regularization, $\gamma$ is the Euler constant.

\section{Self-energy, vertex  and box diagrams for the scalar boson.}

\subsection{Self-energy diagrams.}

Here they are:

\begin{figure}[tbp] 
  \centering
  \includegraphics[bb=144 601 400 715,width=5.67in,height=2.52in,keepaspectratio]{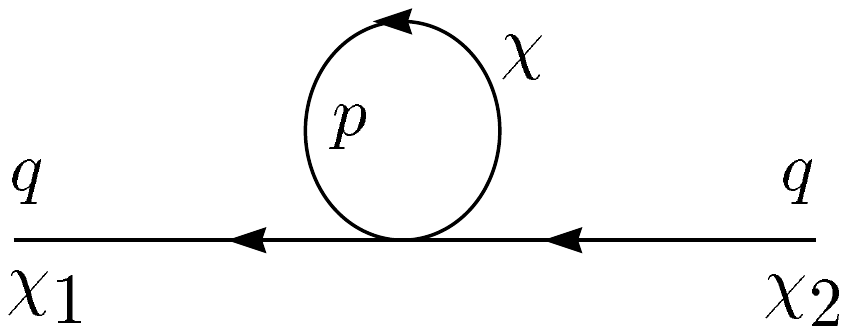}
\caption{}
  \label{fig:FIG-H1}
\end{figure}

\vspace*{3mm}

\begin{equation}
\Pi=\frac{if^{4\chi}}{16\pi^2} M_\chi^2 \left [  2P-1+\log \,{M_\chi^2 \over M_W^2}\right ]  .
\end{equation}

\begin{figure}[tbp] 
  \centering
  \includegraphics[bb=144 601 400 714,width=5.67in,height=2.49in,keepaspectratio]{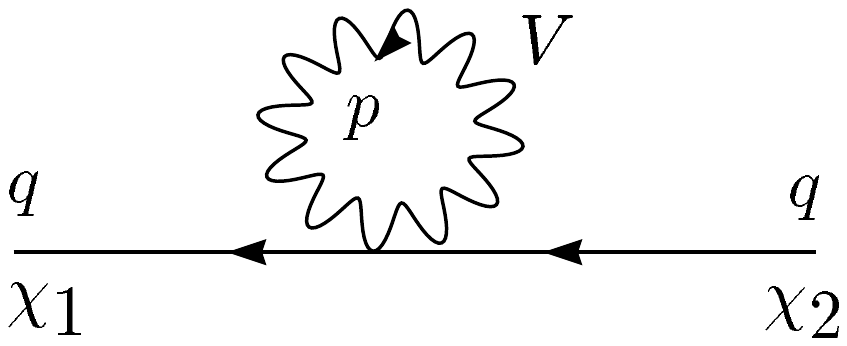}
\caption{}
  \label{fig:FIG-H2}
\end{figure}

\begin{equation}
\Pi=\frac{if^{2V2\chi}}{16\pi^2} M_V^2 \left [  6P-1+3\log{M_V^2 \over M_W^2}\right ]  .
\end{equation}

\begin{figure}[tbp] 
  \centering
  \includegraphics[bb=130 482 386 598,width=5.67in,height=2.57in,keepaspectratio]{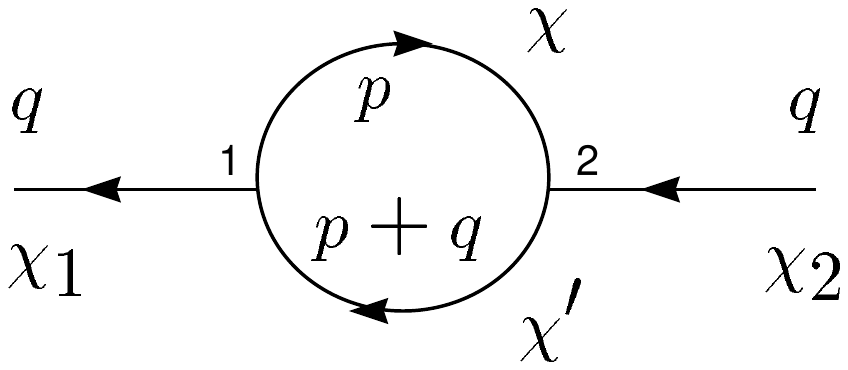}
  \caption{}
  \label{fig:FIG-H3}
\end{figure}

\begin{equation}
\Pi (q^2)=\frac{if_1^{3\chi}f_2^{3\chi}}{16\pi^2}\left [ -2P-I_0(q^2, M_\chi^2, M_{\chi\prime}^2)\right ]  .
\end{equation}

\begin{figure}[tbp] 
  \centering
  \includegraphics[bb=134 488 377 596,width=5.67in,height=2.52in,keepaspectratio]{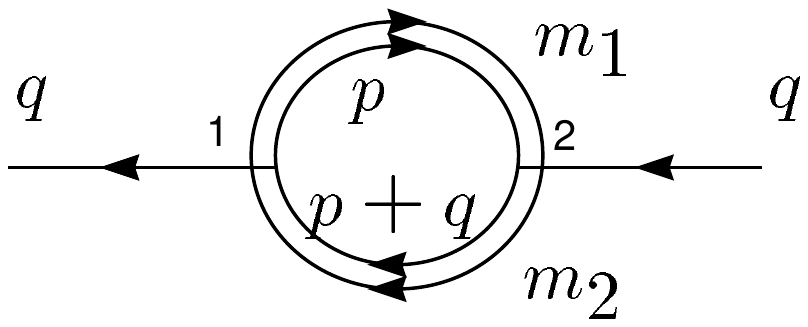}
  \caption{}
  \label{fig:FIG-H4}
\end{figure}

\begin{eqnarray}
\lefteqn{\Pi (q^2)=\frac{if_1^\chi f_2^\chi}{4\pi^2}\left \{\left [ (1-b_1 b_2)\left (q^2+2m_1^2+2m_2^2 \right )
+(1+b_1 b_2)2m_1 m_2\right ]   P+\right.      \nonumber}\\
&+&\left. \left [  {1\over 2}(1-b_1 b_2) \left (q^2+m_1^2+m_2^2 \right )+(1+b_1 b_2)m_1 m_2\right ]   I_0 (q^2, m_1^2, m_2^2)+\right.\\
&+&\left.  {1\over 2} (1-b_1 b_2)\left (m_1^2 \log \,{m_1^2\over M_W^2}+m_2^2 \log \,{m_2^2 \over M_W^2}+{q^2 \over 2}\right )+{1\over 2}(1+b_1 b_2)m_1 m_2\right \}.\nonumber
\end{eqnarray}

\begin{figure}[tbp] 
  \centering
  \includegraphics[bb=130 483 386 602,width=5.67in,height=2.63in,keepaspectratio]{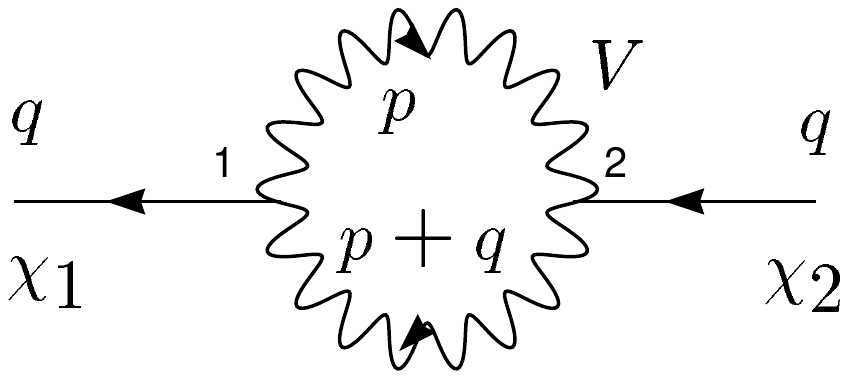}
  \caption{}
  \label{fig:FIG-H5}
\end{figure}

\begin{eqnarray}
\lefteqn{\Pi (q^2)=\frac{if_1^{V\chi} f_2^{V\chi}}{16\pi^2}\left \{\left [ -{q^4\over 2M_V^4}-3{q^2\over M_V^2}-6\right ]   P-\right.}\\
&-&\left.  \left ({q^4\over 4M_V^4}+{q^2\over M_V^2}+3\right )
I_0 (q^2, M_V^2, M_V^2)-{q^2\over 2M_V^2} \log \,{M_V^2 \over M_W^2}+{q^2\over 2M_V^2}-2\right \}.\nonumber
\end{eqnarray}

In the framework of the SM with one Higgs doublet only we obtain
\begin{eqnarray} 
\lefteqn{\Pi^\chi (q^2)=\frac{ig^2}{16\pi^2} M_\chi^2 \left \{\left  [-{3\over 4} {q^4 \over M_W^2 M_\chi^2}-3{q^2 \over M_\chi^2}-{3\over 2R}{q^2 \over M_\chi^2}+{q^2 \over M_W^2 M_\chi^2} Tr \, m_i^2-\right.      \right. }\nonumber\\ 
&-&\left. \left.  3r_W-9r_W^{-1}-{9 \over 2R} r_Z^{-1}+{6\over M_W^2 M_\chi^2} Tr \,m_i^4\right ] P +tadpoles+{3q^4\over 4M_W^2 M_\chi^2}+\right. \nonumber\\ 
&+&\left.  \left ({5\over 2}+{5\over 4R}\right ){q^2 \over M_\chi^2}+{21 \over 8}r_W+{9\over 2}r_W^{-1
}+ {9\over 4R}r_Z^{-1}-{3\over 2}r_W \log \, r_W+\left  ( {q^4 \over 8M_W^4}+\right.      \right.      \nonumber\\
&+&\left. \left.  {3\over 4R}{q^2\over M_W^2}+{9 \over 4R^2}\right  ) r_W^{-1} \log \, R-
{3q^2 \over 4M_W^2 M_\chi^2} Tr \, m_i^2+\right. \nonumber\\
&+&\left.  {q^2 \over 2M_W^2 M_\chi^2} Tr \, m_i^2 \log \,{m_i^2 \over M_W^2}-
 {7 \over 2M_W^2 M_\chi^2} Tr \,m_i^4+{3 \over M_W^2 M_\chi^2} Tr \, m_i^4 \log \,{m_i^2 \over M_W^2}- \right.\nonumber\\
&-& \left .\left ( {q^2 \over 8M_W^2}+{1\over 2}+{3M_W^2 \over 2q^2}\right  ) {1\over M_\chi^2}
 L(q^2, M_W^2, M_W^2) - \right.\nonumber\\
&-&\left. \left ( {q^2 \over 16M_Z^2}+ {1\over 4}+{3M_Z^2 \over 4q^2} \right  ) {1\over R}{1\over M_\chi^2} L(q^2, M_Z^2, M_Z^2)- 
\right.\\ 
&-& \left. {9\over 16} r_W {1\over q^2} L(q^2, M_\chi^2,  M_\chi^2)- {1\over 4M_W^2 M_\chi^2} Tr \, m_i
^2 L(q^2, m_i^2, m_i^2) + \right.      \nonumber\\ &+& \left.  {1\over M_W^2 M_\chi^2}{1\over q^2} Tr \, m_i^4 L(q^2, m_i^2, m_i^2)\right 
\} .\nonumber
\end{eqnarray}  
The corresponding counterterms are
\begin{eqnarray}
\lefteqn{\frac{\delta M_\chi^2}{M_\chi^2}=Z_{M_\chi}-Z_\chi=\frac{ig^2}{16\pi^2}\left \{\left [
3+{3\over 2R}-{15\over 4}r_W- 9r_W^{-1}-{9\over 2R}r_Z^{-1}-{1\over M_W^2} Tr\, m_i^2 +\right.      \right.      \nonumber}\\
&+&\left. \left.  {6\over M_W^2 M_\chi^2} Tr\, m_i^4\right ]   P+ \frac{\Pi^\chi (tadpoles)}{M_\chi^2} -{5\over 2}-{5\over 4R}+{27\over 8}r_W+{9\over 2}r_W^{-1}+{9\over 4R} r_Z^{-1}-\right.      \nonumber\\
&-&\left.  {3\over 2}r_W \log \, r_W + \left ( {1\over 8}r_W-{3\over 4R}+{9\over 4R} r_Z^{-1}\right ) \log \,R+{3\over 4 M_W^2} Tr \, m_i^2-\right.     \nonumber\\
&-&\left.  {1\over 2M_W^2} Tr\, m_i^2 \log \,{m_i^2\over M_W^2} - {7\over 2M_W^2 M_\chi^2} Tr\, m_i^4 +{3\over M_W^2 M_\chi^2} Tr\, m_i^4 \log \,{m_i^2\over M_W^2}+ \right.      \nonumber\\
&+&\left.  \left ({r_W\over 8}-{1\over 2} + {3\over 2} r_W^{-1}\right ){1\over M_\chi^2} L(-M_\chi^2, M_W^2, M_W^2)+\right.      \nonumber\\
&+&\left.  \left ({r_Z\over 16}-{1\over 4} + {3\over 4}r_Z^{-1}\right ){1\over R}{1\over M_\chi^2} L(-M_\chi^2, M_Z^2, M_Z^2)+\right.\\
&+&\left. {9\over 16M_W^2} L(-M_\chi^2, M_\chi^2, M_\chi^2)+ \right.\nonumber\\
&+&\left. {1\over 4M_W^2 M_\chi^2} Tr\, m_i^2 L(-M_\chi^2, m_i^2, m_i^2)-{1\over M_W^2 M_\chi^4} Tr\, m_i^4 L(-M_\chi^2, m_i^2, m_i^2)\right \}\nonumber
\end{eqnarray}
and
\begin{eqnarray}
\lefteqn{Z_\chi - 1=\frac{ig^2}{16\pi^2} \left \{ \left [ -3 - {3\over 2R}+{3\over 2} r_W+{1\over M_W^2} Tr\, m_i^2\right ]   P+{3\over 2}+{3\over 4R}+3r_W^{-1}+\right.\nonumber}\\
&+& \left. {3\over 2R} r_Z^{-1}+
\left ({3\over 4R}-{1\over 4}r_W \right ) \log \,R - {1\over 4M_W^2} Tr\, m_i^2+{1\over 2M_W^2} Tr\, m_i^2 \log \,{m_i^2 \over M_W^2}-\right. \nonumber\\
&-&\left. {2\over M_W^2 M_\chi^2} Tr\, m_i^4+\left ({1\over 4}-{1\over 4}r_W-{3\over r_W (r_W -4)}\right ){1\over M_\chi^2} L(-M_\chi^2, M_W^2, M_W^2)+\right.      \nonumber\\
&+&\left.  \left  ({1\over 8}-{1\over 8} r_Z-{3\over 2r_Z (r_Z-4)}\right  ) {1\over R} {1\over M_\chi^2} L(-M_\chi^2, M_Z^2, M_Z^2)+\right.\nonumber\\
&+& \left. {3\over 8}{1\over M_W^2} L(-M_\chi^2, M_\chi^2, M_\chi^2)-\right.\\
&-&\left.   {1\over 4M_W^2 M_\chi^2} Tr\, m_i^2 L(-M_\chi^2, m_i^2, m_i^2)-
{1\over 2M_W^2 M_\chi^4} Tr\, m_i^4 L(-M_\chi^2, m_i^2, m_i^2)\right \}\nonumber
\end{eqnarray}

Consequently,
\begin{eqnarray}
\lefteqn{\Pi^{ren}(q^2)=\Pi^\chi (q^2)-\delta M_\chi^2 - (Z_\chi - 1) (q^2 +M_\chi^2)=\frac{ig^2}{16\pi^2} M_\chi^2 \times}\nonumber\\
&\times& \left \{ \left[ -{3\over 4}{q^4 \over M_W^2 M_\chi^2}-{3\over 4}r_W-      
{3\over 2}{q^2 \over M_W^2}\right ]   P+{3q^4 \over 4M_W^2 M_\chi^2}+{q^4 \over 8M_W^2 M_\chi^2}\log \,R +\right.\nonumber\\
&+&\left.{q^2\over M_\chi^2} \left (1+{1\over 2R}-3r_W^{-1}-{3\over 2R}r_Z^{-1}\right )+
{q^2\over 4M_W^2}\log \,R+{1\over 8}r_W \log \,R+\right.\nonumber\\
&+&\left. 1+{1\over 2R}- {3\over 4}r_W-3r_W^{-1}-{3\over 2R}r_Z^{-1}-
\left ({q^2\over M_\chi^2}+1\right ){1\over 2M_W^2} Tr\, m_i^2+\right.\nonumber\\
&+&\left.\left ({q^2\over M_\chi^2}+1\right ){2\over M_W^2 M_\chi^2} Tr\, m_i^4 - \left ({q^2\over 8 M_W^2}+{1\over 2}+{3M_W^2\over 2q^2}\right )\times\right. \nonumber\\
&\times&\left.  {1\over M_\chi^2} L(q^2, M_W^2, M_W^2)-\left ({q^2\over 16 M_Z^2}+{1\over 4}+{3M_Z^2\over 4q^2}\right ){1\over R}{1\over M_\chi^2}L(q^2, M_Z^2, M_Z^2)-\right.      \nonumber\\
&-&\left.  {9\over 16 q^2} r_W L(q^2, M_\chi^2, M_\chi^2)+{1\over 4M_W^2 M_\chi^2} Tr\, m_i^2 L(q^2, m_i^2, m_i^2)+\right.      \nonumber\\
&+&\left.  {1\over M_W^2 M_\chi^2} {1\over q^2}Tr\, m_i^4 L(q^2, m_i^2, m_i^2)+ \left (-{q^2\over 4M_\chi^2}+{q^2\over 4M_W^2} - {3q^2\over M_\chi^2 r_W (r_W-4)}+\right.       \right.      \nonumber\\
&+&\left. \left. {1\over 4}+{1\over 8} r_W-{3\over 2}r_W^{-1}+{3\over r_W(r_W-4)}\right ){1\over M_\chi^2} L(-M_\chi^2, M_W^2, M_W^2)+\right. \nonumber\\
&+&\left.  \left (-{q^2\over 8M_\chi^2}+{q^2\over 8M_Z^2}+{3q^2\over M_\chi^2 r_Z (r_Z-4)}+{1\over 8}+
 {1\over 16}r_Z-{3\over 4}r_Z^{-1}+ \right.\right.\nonumber\\
&+&\left.\left. {3\over 2r_Z(r_Z-4)}\right )
 {1\over R}{1\over M_\chi^2}L(-M_\chi^2, M_Z^2, M_Z^2)- \right.\nonumber\\
&-&\left. \left ({3q^2\over 8M_W^2}+{15\over 16} r_W\right ){1\over M_\chi^2} L(-M_\chi^2, M_\chi^2, M_\chi^2) +\right.      \nonumber\\
&+&\left.  {q^2\over 4M_W^2 M_\chi^4} Tr\, m_i^2 L(-M_\chi^2, m_i^2, m_i^2)+
{1\over 2 M_W^2 M_\chi^4}\left (3+{q^2\over M_\chi^2}\right ) \times \right.   \nonumber\\ &\times&\left.  Tr\, m_i^4 L(-M_\chi^2, m_i^2, m_i^2) \right\}
\end{eqnarray}
In  this Section and in what follows $r_W=M_\chi^2/M_W^2$, $r_Z=M_\chi^2/M_Z^2$,
$b_{1,2}$ are constants defined by the strength of the Higgs -fermion pseudoscalar interaction.
\footnote{In the Standard Model they are equal to zero.}
The form of the integral $I_0 (q^2, M_1^2, M_2^2)$ is given in {\it Appendix~A}.
\begin{figure}[tbp] 
  \centering
\includegraphics[scale=0.6]{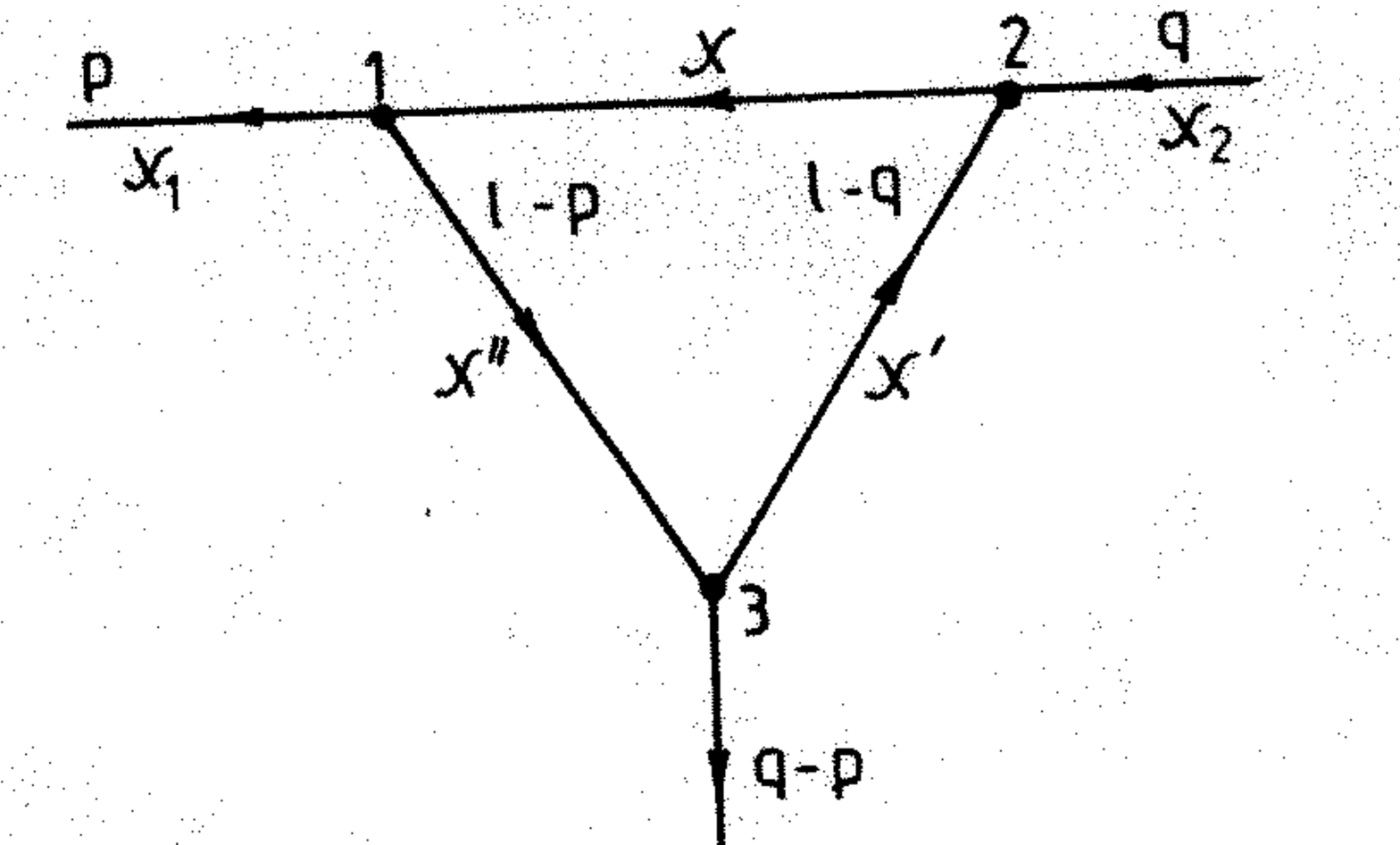} 
 \caption{}
 \label{fig:FIG-H6}
\end{figure}

\subsection{Vertex diagrams.}

The technique of the calculation of the diagrams shown below is very much alike to that suggested in paper~\cite{14}.
\begin{equation}
\Gamma (p^2, q^2, (p-q)^2)=\frac{i f_1^{3\chi}f_2^{3\chi}f_3^{3\chi}}{16\pi^2} I_1 (q^2, (p-q)^2, p^2, M_\chi^2, M_{\chi^{\prime}}^2, M_{\chi^{\prime\prime}}^2 ).
\end{equation}
and a similar diagram with the opposite lepton current direction give
\begin{figure}[tbp] 
 \centering
\includegraphics[scale=0.6]{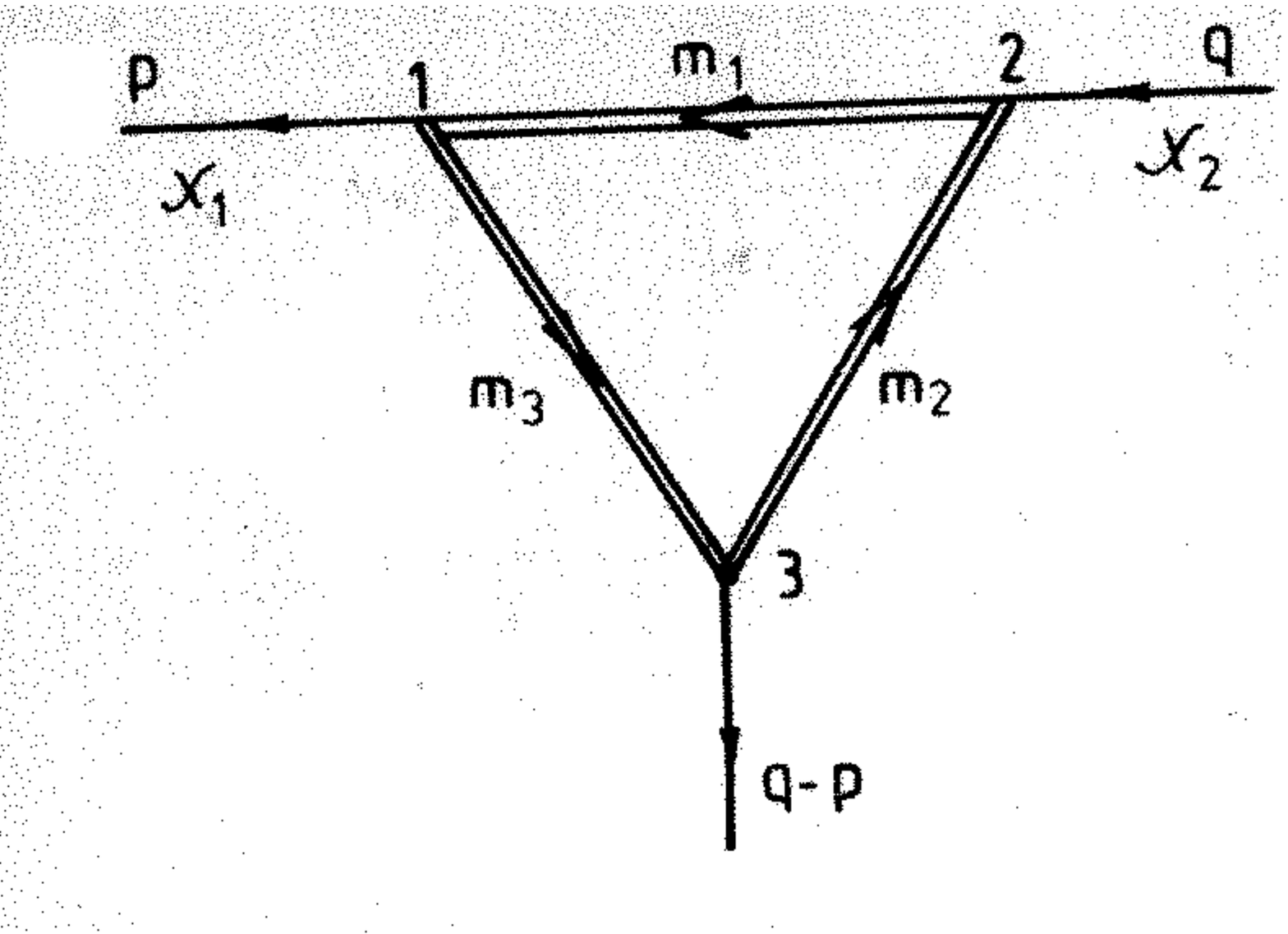} 
 \caption{}
  \label{fig:FIG-H7}
\end{figure}

\begin{eqnarray}
\lefteqn{\Gamma (p^2, q^2, (p-q)^2)=\frac{i f_1^\chi f_2^\chi f_3^\chi}{4\pi^2}\left \{ \left [ -2m_1 B_1 -2m_2 B_2 - 2m_3 B_3\right ] P-\right.      \nonumber}\\
&-&\left. \frac{m_1 B_1 + m_2 B_2 +m_3 B_3}{2} - \frac{m_1 B_1 +m_2 B_2}{2} I_0 (q^2, m_1^2, m_2^2) - \right.      \nonumber\\
&-&\left.  \frac{m_1 B_1 +m_3 B_3}{2} I_0 (p^2, m_1^2, m_3^2) -\frac{m_2 B_2 +m_3 B_3}{2} I_0 ((p-q)^2, m_2^2, m_3^2) -   \right.       \nonumber\\
&-&\left.  {1\over 2} \left [  \left  ( m_1 (p-q)^2 +m_1 m_2^2 +m_1 m_3^2 \right ) B_1 + \right.\right.\nonumber\\
&+&\left.\left. \left  (m_2 p^2 +m_2 m_1^2 +m_2 m_3^2 \right  ) B_2 +
 \left  (m_3 q^2 +m_3 m_1^2 +m_3 m_2^2\right  ) B_3 + \right.\right.\nonumber\\
&+&\left. \left.  2m_1 m_2 m_3 B_4\right ]
I_1 (q^2, (p-q)^2, p^2, m_1^2, m_2^2, m_3^2)\right \}.\nonumber
\end{eqnarray}
where
\begin{eqnarray}
B_1&=&1+b_1 b_2 -b_1 b_3 -b_2 b_3,\nonumber\\
B_2&=& 1-b_1 b_2 -b_1 b_3 +b_2 b_3,\nonumber\\
B_3&=&1- b_1 b_2 +b_1 b_3 -b_2 b_3,\nonumber\\
B_4 &=&1+b_1 b_2+ b_1 b_3 +b_2 b_3.
\end{eqnarray}

\begin{figure}[tbp] 
  \centering
\includegraphics[scale=0.6]{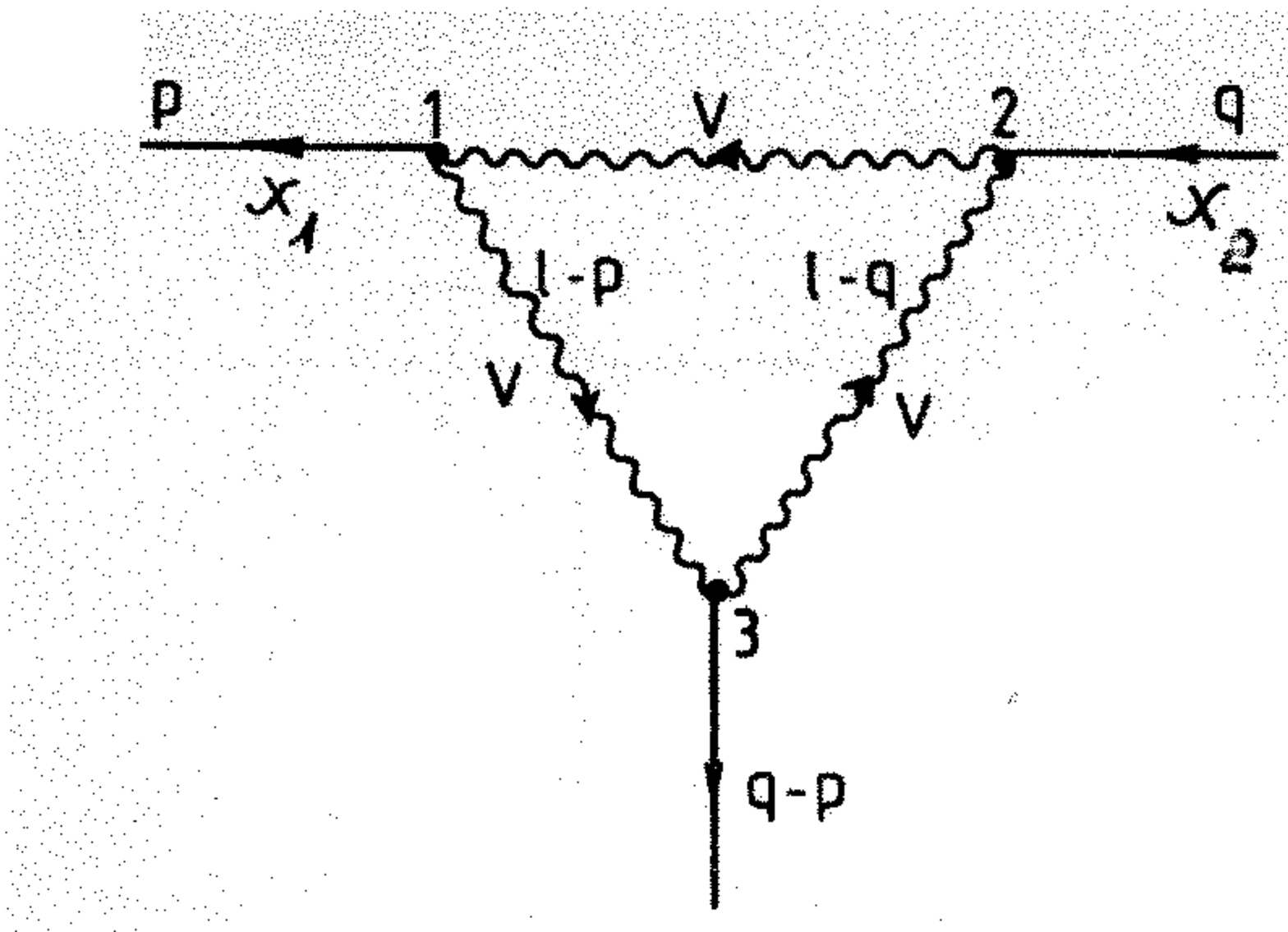} 
  \caption{}
  \label{fig:FIG-H8}
\end{figure}
and a similar diagram with the opposite loop current direction
\begin{eqnarray}
\lefteqn{\Gamma (p^2, q^2, (p-q)^2) = \frac{if_1^{V\chi} f_2^{V\chi} f_3^{V\chi}}{16\pi^2} \left \{
\left [  -{3\over 4}\frac{p^2 +q^2 +(p-q)^2}{M_V^4} - \right.\right.}\nonumber\\
&-&\left.\left. {1\over 8} \frac{(p^2 +q^2 + (p-q)^2)^2}{M_V^6} \right ]   P +
 \frac{p^2 +q^2 +(p-q)^2}{4M_V^4} \left  (1- \log \,{M_V^2 \over M_W^2}\right )-\right.\nonumber\\
&-& \left.\left [ {1\over 4M_V^4} \left (\frac{p^2 +q^2 - (p-q)^2}{2}+p^2 -q^2 \right ) +
\frac{p^2 (p^2 +q^2 +(p-q)^2)}{16 M_V^6}\right ]   \times \right.\nonumber\\
&\times& \left. I_0 (p^2, M_V^2, M_V^2)-
 \left [ {1\over 4M_V^4} \left ({p^2 +q^2 -(p-q)^2 \over 2}+q^2 -p^2 \right ) +\right.\right.   \nonumber\\
&+&\left. \left. \frac{q^2 (p^2 +q^2 +(p-q)^2)}{16 M_V^6}\right ]   I_0 (q^2, M_V^2, M_V^2)+ \left [  {1\over 4M_V^4}\left ({p^2 +q^2 - (p-q)^2 \over 2}- \right. \right.\right.   \nonumber\\
&-&\left.  \left. \left. (p-q)^2\right ) - \frac{(p-q)^2 (p^2+q^2 +(p-q)^2)}{16M_V^6}\right ]   I_0 ((p-q)^2, M_V^2, M_V^2)+ \right.\nonumber\\
&+&\left. \left [  3+\frac{p^2 +q^2 +(p-q)^2}{2M_V^2}+\right.\right.\nonumber\\
&+&\left.\left.
 \frac{p^4 +q^4-p^2 q^2 +(p-q)^4 -p^2 (p-q)^2 -q^2 (p-q)^2}{4M_V^4} -\right.\right.\nonumber\\
&-& \left.\left. \frac{p^2 q^2 (p-q)^2}{8M_V^6} \right  ]   I_1 (q^2, (p-q)^2, p^2, M_V^2, M_V^2, M_V^2) \right \}.
\end{eqnarray}
The form of the integral $I_1$ is also given in {\it Appendix A}.

The diagrams can be easily derived from the self-energy diagrams (see preceding Subsection).  As a sum, we get
\begin{figure}[tbp] 
  \centering
  \includegraphics[bb=-2 148 520 727,width=5.67in,height=6.28in,keepaspectratio]{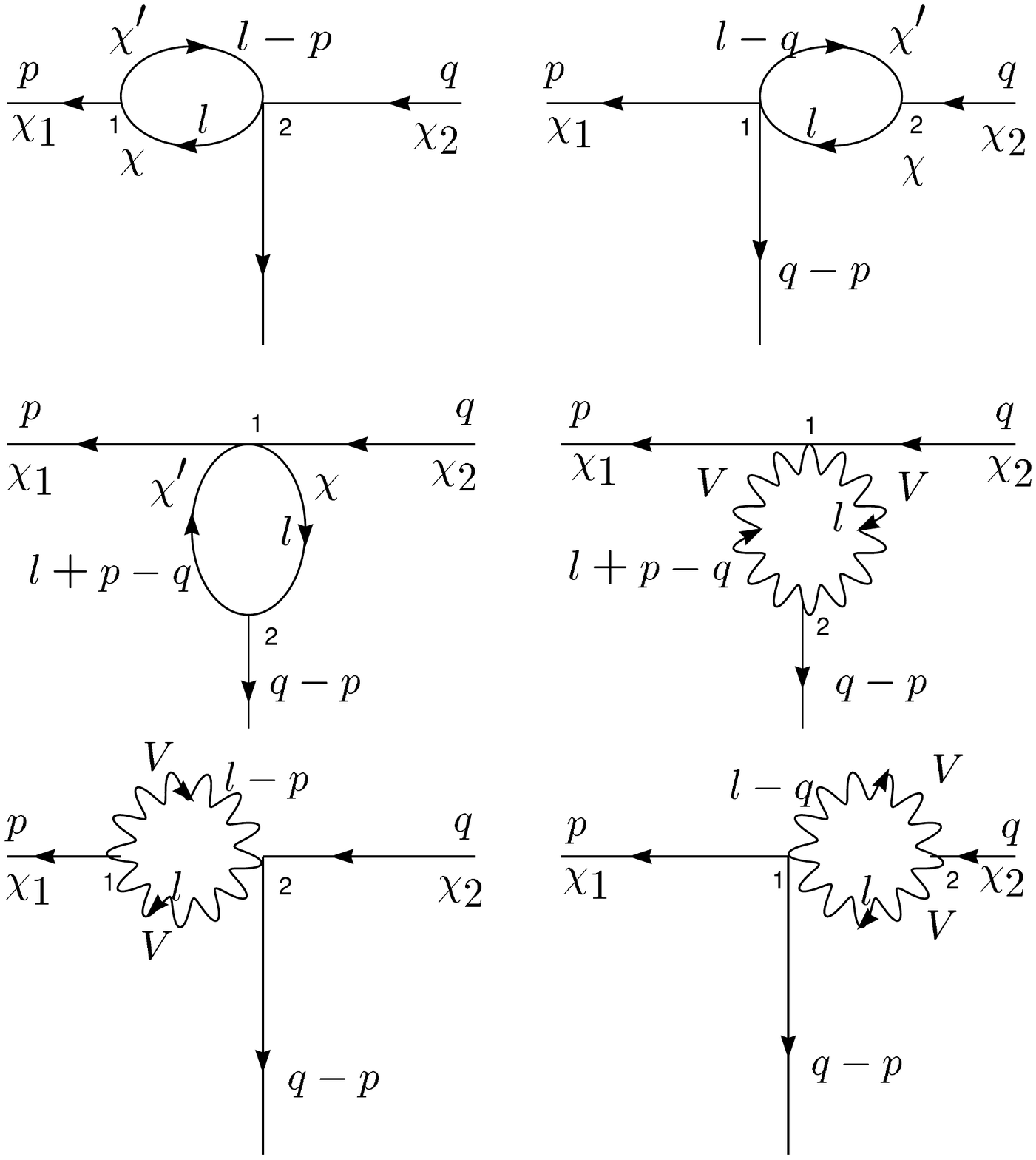}
  \caption{}
  \label{fig:FIG-H9}
\end{figure}

\begin{eqnarray}
\lefteqn{\Gamma (p^2, q^2, (p-q)^2) = \frac{ig^3}{16\pi^2}\frac{M_\chi^2}{M_W}
\left  \{ \left  [  {3\over 4} \frac{p^2 q^2 +p^2 (p-q)^2 +q^2 (p-q)^2}
{M_W^2 M_\chi^2} - {27 \over 8} r_W - \right.   \right.  }\nonumber\\
&-&\left. \left.  9r_W^{-1} -
{9\over 2R} r_Z^{-1} +{6\over M_W^2 M_\chi^2} Tr \, m_i^6\right  ]   P+
\Gamma^{3\chi} (tadpoles) -\frac{p^2 +q^2 +(p-q)^2}{4M_\chi^2}+\right. \nonumber\\
&+&\left.  \frac{p^2 +q^2 +(p-q)^2}{8M_\chi^2}{1\over R} - {3\over 4}
\frac{p^2 q^2 +p^2 (p-q)^2 +q^2 (p-q)^2}{M_W^2 M_\chi^2} -\right.  \nonumber\\
&-&\left.  \frac{p^2 q^2 +p^2 (p-q)^2 +q^2 (p-q)^2}{8M_W^2 M_\chi^2} \log R + 
  {27\over 8} r_W -{27\over 16}r_W \log r_W + 6r_W^{-1}+ \right.\nonumber\\
&+&\left. 3 {1\over R} r_Z^{-1} +
+{9\over 4} r_Z^{-1} {1\over R} \log R - {9\over 2 M_W^2 M_\chi^2} Tr\, m_i^4+
\frac{3}{M_W^2 M_\chi^2} Tr \ m_i^4 \log {m_i^2 \over M_W^2} -\right.\nonumber\\
&+&\left.
\left [ -{3 \over 2}r_W^{-1}-\frac{q^2+(p-q)^2-p^2}{4M_\chi^2}+
\frac{p^2(q^2+(p-q)^2 )}{8M_W^2 M_\chi^2}\right ]   \times \right. \nonumber\\ 
&\times& \left.  {1\over 2p^2} L(p^2,M_W^2, M_W^2)+
\left [  -{3 \over 2} r_W^{-1}-\frac{p^2+(p-q)^2-q^2}{4M_\chi^2}+ \frac{q^2(p^2+(p-q)^2 )}{8M_W^2 M_\chi^2} \right  ] \times \right. \nonumber\\ 
&\times& \left. {1\over 2q^2} L(q^2,M_W^2, M_W^2)
+   \left  [  - {3 \over 2} r_W^{-1} - \frac{p^2+q^2-(p-q)^2-p^2}{4M_\chi^2}+\right.\right.\nonumber\\
&+&\left.\left. \frac{(p-q)^2 (p^2+q^2 )}{8M_W^2 M_\chi^2 } \right  ]
 {1\over (p-q)^2 } L((p-q)^2, M_W^2, M_W^2 )+ \right.\nonumber\\
&+&\left. \left  [  - {3\over 4R} r_Z^{-1} - \frac{q^2+(p-q)^2-p^2}{8M_\chi^2} {1\over R} + \frac{p^2 (q^2+(p-q)^2 )}{16M_W^2  M_\chi^2 } \right  ]  {1\over 2p^2} L(p^2, M_Z^2, M_Z^2)+\right.    \nonumber\\
&+&\left.  \left  [  -{3\over 4R}r_Z^{-1}-\frac{p^2+(p-q)^2-q^2}{8M_\chi^2}{1\over R}+
\frac{q^2(p^2+(p-q)^2 )}{16M_W^2  M_\chi^2}\right  ] {1\over 2q^2} L(q^2, M_Z^2, M_Z^2)+
\right.    \nonumber\\
&+&\left.  \left  [  -{3\over 4R}r_Z^{-1}-\frac{p^2+q^2-(p-q)^2-p^2}{8M_\chi^2}{1\over R}+\right.\right.\nonumber\\
&+&\left.\left.\frac{(p-q)^2(p^2+q^2 )}{16M_W^2  M_\chi^2}\right  ] {1\over 2(p-q)^2} L((p-q)^2, M_Z^2, M_Z^2)-\right. \nonumber\\ 
&-& \left.  { 9\over 16} r_W \left [ {1\over 2p^2} L (p^2, M_\chi^2, M_\chi^2) + {1\over 2q^2} L(q^2, M_\chi^2, M_\chi^2) +\right.\right.\nonumber\\
&+&\left.\left.
{1\over  2 (p-q)^2} L((p-q)^2, M_\chi^2, M_\chi^2) \right ] +\right. \nonumber\\ 
&+&\left. {1\over M_W^2M_\chi^2} Tr\, m_i^4\left [ {1\over 2p^2} L(p^2,M_\chi^2,M_\chi^2)+ {1\over2q^2}L(q^2,M_\chi^2,M_\chi^2)+
\right.\right.\nonumber\\
&+&\left.\left.{1\over (p-q)^2} L((p-q)^2,M_\chi^2, M_\chi^2)\right  ] - 
{27 \over 8} r_W M_\chi^2  I_1 (q^2, (p-q)^2, p^2, M_\chi^2,
M_\chi^2, M_\chi^2) -\right.   \nonumber\\ &-&\left. \left [ 6r_W^{-1}M_W^2 + (p^2 +q^2 +(p-q)^2) r_W^{-1} + \right.\right.\nonumber\\
&+&\left. \left.\frac{p^4 +q^4 +(p-q)^4 -p^2 q^2 -p^2 (p-q)^2 -q^2 (p-q)^2}{2M_\chi^2} - \right. \right. \nonumber\\ 
&-&\left. \left. \frac{p^2 q^2 (p-q)^2}{4M_W^2 M_\chi^2}\right ]  I_1 (q^2, (p-q)^2, p^2, M_W^2, M_W^2, M_W^2)-\right.    \nonumber\\
&-&\left.  \left  [ {3\over R} r_Z^{-1}M_Z^2 + \frac{p^2 +q^2 +(p-q)^2}{2R} r_Z^{-1}+\right.\right.\nonumber\\
&+&\left.\left.\frac{p^4 +q^4  + (p-q)^4-p^2 q^2 -p^2 (p-q)^2 -q^2 (p-q)^2}{4R M_\chi^2}- \right. \right. \nonumber\\ &-&\left. \left. \frac{p^2 q^2 (p-q)^2}{8M_W^2 M_\chi^2} \right ] I_1 (q^2, (p-q)^2, p^2, M_Z^2, M_Z^2, M_Z^2)+ \right. \nonumber\\ 
&+&\left. \frac{p^2 +q^2 + (p-q)^2}{2M_W^2 M_\chi^2} Tr m_i^4 I_1 (q^2, (p-q)^2, p^2, m_i^2, m_i^2, m_i^2) +\right. \nonumber\\ &+&\left.  {4\over M_W^2 M_\chi^2} Tr m_i^6 I_1 (q^2, (p-q)^2, p^2, m_i^2,
m_i^2, m_i^2)\right  \}.
\end{eqnarray}  
The appropriate counterterm is 
\begin{equation}
\Gamma^{c.t}=-{3g M_\chi^2\over 2M_W}\left [ {3\over 2}(Z_\chi - 1)+{\delta g\over g}+{\delta M_\chi^2\over M_\chi^2}-{\delta M_W^2\over 2M_W^2}\right ] .
\end{equation} 
As found in~\cite{8} 
\begin{equation}
{\delta g\over g}=Z_A^{-1/2}\left (1-\frac{\delta R}{1-R}\right )^{-1/2}-1 \simeq {1\over 2}\left [ \frac{\delta R}{1-R}- (Z_A -1)\right ]  ,
\end{equation} 
\begin{equation}
\frac{\delta R}{R}=\frac{Z_{M_W}Z_W^{-1}}{Z_{M_Z}Z_Z^{-1}}-1 \simeq \frac{\delta M_W^2}{M_W^2} - \frac{\delta M_Z^2}{M_Z^2},
\end{equation} 
where 
\begin{eqnarray}
\lefteqn{\frac{\delta M_W^2}{M_W^2}=Z_{M_W}-Z_W=
\frac{ig^2}{16\pi^2}\left \{\left [  {34\over 3} -{3\over 2R}-{1\over 3} N_f
+{1\over M_W^2} Tr\, m_i^2\right ]   P- 19+{14\over 9}+\right.   }\nonumber\\
&+&\left.   {23\over 12 R}+ {1\over 12R^2} -{1\over 2} r_W +{1\over 12} r_W^2 +
(-{7\over 2}+ {7 \over 12 R} +{1\over 24 R^2}){1\over R} \log R +\right. \nonumber\\
&+& \left.  (-{3\over 4} +{1\over 4}r_W-{1\over 24}r_W^2) r_W \log r_W
+{7\over 36} N_f -{1\over 12 M_W^2} Tr m_i^2 -{1\over 6M_W^4}Tr m_i^4+\right. \nonumber\\
&+& \left. \sum_{i,j}^{N_f/2} \left [ {m_i^2 m_j^2\over 3M_W^4}-({1\over 3}-\frac{m_i^2 +m_j^2}{2M_W^2}\times
\log {m_i m_j}{M_W^2} +\frac{(m_i^2 -m_j^2)^3}{12M_W^6}\log {m_i^2}{m_j^2}-\right.   \right.   \nonumber\\
&-&\left. \left.  ({1\over 6} -\frac{m_i^2 +m_j^2}{12M_W^2}- \frac{m_i^2 -m_j^2)^2}{12M_W^4}){1\over M_W^2}
L(-M_W^2, m_i^2, m_j^2)\right ]  K_{ij} K_{ij}^+ +\right.   \nonumber\\
&-&\left.  ({1\over 2}-{r_W\over 6}+{r_W^2\over 24}){1\over M_W^2} L(-M_W^2,
M_W^2, M_\chi^2) +(-{17\over 6}-2R+{2\over 3R}+{1\over 24 R^2})\times\right. \nonumber\\
&\times&\left. {1\over M_W^2}
L(-M_W^2, M_W^2, M_Z^2)\right \},
\end{eqnarray} 
and 
\begin{eqnarray}
\lefteqn{\frac{\delta M_Z^2}{M_Z^2}=Z_{M_Z}-Z_Z=
\frac{ig^2}{16\pi^2}\left \{\left [ -{7\over 3} +14R-{11\over 6R}-\right. \right.    \nonumber}\\
&-&\left. \left. {1 \over 3}\left ( 2-{1\over 3} \right ) N_f -{8\over 3} \frac{(1-R)^2}{R}
Tr\, Q_i^2 +{1\over M_W^2} Tr\, m_i^2 \right ]  P+\right.\nonumber\\
&+&\left.{35 \over 18} -{34 \over 3} R -
8R^2+\right.   \nonumber\\
&-&\left. {35 \over 18R} - {1\over 2} r_W +{1\over 12} r_W r_Z +{5\over 6R} \log R +
\left ( -{3\over 4} +{1\over 4}r_Z -{1\over 24} r_Z^2 \right ) r_W \log r_Z+\right.   \nonumber\\
&+&\left.  \left ( {7\over 36} N_f -{14\over 9} Tr \, Q_i^2\right )\left (2 -{1\over R}\right )
+{14\over 9} R Tr\, Q_i^2+ \right.   \nonumber\\
&+& \left.  {1\over M_W^2} Tr\, \left [  \left ( -{1\over 12} -{8\over 3}
\vert Q_i\vert +{16\over 3}Q_i^2\right )m_i^2 + \left ( {8\over 3} \vert Q_i
\vert -{32\over 3} Q_i^2\right ) R m_i^2 +\right.
\right.   \nonumber\\
&+&\left. \left. {16\over 3} Q_i^2 R^2 m_i^2\right ] + {1\over M_W^2} Tr\, \left [  {1\over 2} m_i^2 -{1\over 3}M_Z^2
\left ( {1\over 2} -2 \vert Q_i \vert + 4Q_i^2\right )- \right.\right.\nonumber\\
&-&\left.\left. {1\over 3}M_W^2
\left (2\vert Q_i \vert -8Q_i^2\right ) -{4M_W^2 \over 3M_Z^2} Q_i^2 \right ] \log {m_i^2 \over M_W^2}+\right.   \nonumber\\
&+&\left.  \left ({1\over 24} +{2\over 3} R -{17\over 6} R^2 -2R^3\right )
{1\over M_W^2} L(-M_Z^2, M_W^2, M_W^2) +\right.\nonumber\\
&+&\left.
\left ( {1\over 2} -{1\over 6} r_Z +{1\over 24} r_Z^2\right )
{1\over M_W^2} L(-M_Z^2, M_Z^2, M_\chi^2) +\right.\nonumber\\
&+& \left.  Tr\, \left [  \left ( {1\over 12}-{1\over 3} \vert Q_i\vert +{2\over 3} Q_i^2\right ) +
\left ( {1\over 3}\vert Q_i \vert -{4\over 3} Q_i^2 \right ) R +{2\over 3}Q_i^2 R^2-\right.   \right.   \nonumber\\
&-&\left. \left.  \left ({1\over 12} +{2\over 3} \vert Q_i\vert -
{4\over 3} Q_i^2 \right )
{m_i^2\over M_Z^2} + \left ({2\over 3} \vert Q_i\vert -{8\over 3} Q_i^2 \right ) R {m_i^2 \over M_Z^2}+
\right.\right.\nonumber\\
&+&\left.\left.
{4\over 3}Q_i^2 R^2 {m_i^2\over M_Z^2}\right ]     {1\over M_W^2} L(-M_Z^2, m_i^2, m_i^2)
\right \}
\end{eqnarray} 
Finally,\footnote{Some difference of the renormalization constants from those given in~\cite{8} appears firstly due to the approximation $s, t, u, M_V^2, M_\chi^2 \gg m_f^2$ done there (and not used in our consideration) and, secondly, due to the different definition of the $f(d)$ function (used in the trace calculations, see {\it Appendix A}), that is known to not have influence the physical quantities.} 
\begin{equation}
Z_A - 1=\frac{ie^2}{16\pi^2}\left \{\left [  -14+{8\over 3}Tr\, Q_i^2\right ] P+{2\over 3} \left (1+Tr\,Q_i^2\right )+{4\over 3}Tr\, Q_i^2 \log \,{m_i^2\over m_W^2}\right \}.
\end{equation} 
Thus,
\begin{eqnarray}
\lefteqn{\frac{\delta g}{g}=\frac{ig^2}{16\pi^2}\left \{ \left [  {43\over 6}-{1\over 6}N_f \right ]   P-{1\over 3}(1-R)\left (1+Tr\, Q_i^2+\right. \right. \nonumber}\\
&+&\left. \left.  2 Tr\, Q_i^2 \log \,{m_i^2 \over M_W^2}\right )+{1\over 2}\frac{R}{1-R}\left [ W(-1) - Z(-1)\right ]  \right \},
\end{eqnarray} 
where $W(-1)$ and $Z(-1)$ are the finite parts of the counterterms $\delta M_W^2/M_W^2$ and $\delta M_Z^2/M_Z^2$, respectively.  Then, 
\begin{eqnarray}
\lefteqn{\Gamma^{c.t.} (p^2, q^2, (p-q)^2)=-\frac{ig^3}{16\pi^2}\frac{3M_\chi^2}{2M_W}
\left \{ \left [  - 
{3\over 2} r_W -9r_W^{-1} - \right.\right.}\nonumber\\
&-&\left. \left. {9\over 2R} r_Z^{-1}+{6\over M_W^2 M_\chi^2} Tr\,m_i^4 \right  ] P+\right.  \nonumber\\
&+&\left.  {1\over 2} \left ({R \over 1-R} - 1\right  ) W(-1) - {1\over 2}{R \over 1-R} Z(-1) +\right.\nonumber\\
&+&\left.\chi (-1) +{3\over 2}\chi^F (-1) -2\Gamma^{3\chi} (tadpoles)-\right.    \nonumber\\
&-&\left.  {1\over 3} (1-R) \left ( 1+ Tr \, Q_i^2 +2 Tr \, Q_i^2 \log \, {m_i^2 \over M_W^2} \right  ) \right \}= \nonumber\\
&=& - {ig^3 \over 16\pi^2} {3M_\chi^2 \over 2M_W} \left  \{ \left  [ -{ 3\over 2}r_W -9r_W^{-1} -
 {9\over 2R} r_Z^{-1} +{6\over M_W^2 M_\chi^2} Tr \, m_i^4 \right   ] P+\right.\nonumber\\
&+&\left.{49 \over 6} - {28 \over 3} R -4R^2 - {25 \over 24R} - {1\over 24R^2} -\right.   \nonumber\\
&-&\left.
\left ( {10 \over 9} - {10\over 9} R\right  ) Tr \, Q_i^2 + {29 \over 8} r _W - {1\over 4} r_W^2 +
{1\over 4} r_W r_Z +  9 r_W^{-1} + {9\over 2R} r_Z^{-1}+\right.   \nonumber\\
&+&\left.   \left (-{9\over 8} +{1\over 8}r_Z-{1\over 8}r_W +{1\over 48} r_W^2 -{1\over 48} r_W r_Z - {1\over 48} r_Z^2\right  ) r_W \log \, r_W+ \right. \nonumber\\
&+&\left.  \left (-{47\over 12}+{7 \over 3R} -{1\over 4R^2} -{1\over 48R^3} +{3\over 8} r_Z - {1\over 8} r_Z^2+ {1\over 48}r_Z^3 \right  ) {1\over 1-R} \log \,R +\right.   \nonumber\\
&+&\left.  \left (-{1\over 4}r_W +{3\over 8R} +{9\over 4R}r_Z^{-1} \right  )
\log \,R +\right.\nonumber\\
&+&\left.
{1\over M_W^2} Tr \, \left [  {5\over 12}m_i^2 + \left ({4\over 3} \vert Q_i \vert - {8\over 3} Q_i^2 \right  ) R m_i^2 +
 {8\over 3} Q_i^2  R^2 m_i^2 \right ]  + \right.\nonumber\\
&+&\left.{1\over M_W^4} Tr \, \left [  {1\over 12} m _i^4 -{1\over 12}{R\over 1-R} m_i^4 -{13\over 2}r_W^{-1} m_i^4 \right  ]+\right.   \nonumber\\
&+& \left.  Tr \, \left [ {m_i^2\over 4M_W^2}\left (1-{R\over 1-R}\right  ) +{1\over 12} {1\over 1-R} -{1\over 3} \vert Q_i \vert + 3 {m_i^4 \over M_W^2 M_\chi^2} \right  ] \log \,{m_i^2 \over M_W^2}+\right.   \nonumber\\
&+&\left.   {R\over 1-R} \left (-{1\over 48}-{1\over 3} R +{17\over 12} R^2 +R^3\right  ) {1\over M_W^2} L (-M_Z^2, M_W^2, M_W^2) +\right.    \nonumber\\
&+&\left.  {R\over 1-R} \left (-{1\over 4}+{1\over 12} r_Z -{1\over 48}r_Z^2\right  ) {1\over M_W^2} L(-M_Z^2, M_Z^2, M_\chi^2) +{1\over 1-R} \left (-{1\over 4} +{1\over 2}R \right.   \right.   \nonumber\\
&+&\left.  \left.  {1\over 12}r_W-{1\over 6}r_Z +{1\over 24}r_W r_Z - {1\over 48} r_W^2\right  ) {1\over M_W^2} L(-M_W^2, M_W^2, M_\chi^2) +\right. \nonumber\\
&+&\left.  {1\over 1-R} \left ({25\over 12} - {11\over 6}R -2R^2 - {7\over 24R} -{1\over 48 R^2} \right  ) {1\over M_W^2} L(-M_W^2, M_W^2, M_Z^2) +\right. \nonumber\\
&+&\left.  \left (-{1\over 4}-{1\over 8}r_W^{-1} +{3\over 2}r_W^{-2} -{9\over 2} {1\over r_W^2 (r_W-4)}\right  ){1\over M_W^2} L(-M_\chi^2, M_W^2, M_W^2)+\right.   \nonumber\\
&+&\left.  \left (-{1\over 8} -{1\over 16}r_Z^{-1} + {3\over 4} r_Z^{-2} -{9\over 4}{1\over r_Z^2 (r_Z-4)}\right  ){1\over M_W^2} L(-M_\chi^2, M_Z^2, M_Z^2) +\right.   \nonumber\\
&+&\left.  {9\over 8M_W^2} L(-M_\chi^2, M_\chi^2, M_\chi^2)- {1\over M_W^2} Tr \, \left [ {1\over 24} {R\over 1-R} -\right.\right. \nonumber\\
&-&\left. \left ({1\over 6}\vert Q_i \vert -{1\over 3} Q_i^2 \right  ) R-{1\over 3} Q_i^2 R^2 -\right.    \nonumber\\
&-&\left. {1\over 24} {R\over 1-R} {m_i^2 \over M_Z^2} - \left ({1\over 3} \vert Q_i \vert -{2\over 3}Q_i^2 \right  ){m_i^2\over M_Z^2} R - {2\over 3}Q_i^2 R^2 {m_i^2\over M_Z^2} \right ]   L(-M_Z^2, m_i^2, m_i^2) - \nonumber\\
&-& {1\over M_W^2}
Tr \, \left [  {m_i^2 \over 8M_\chi^2 }+{7m_i^4\over 8M_\chi^4}\right ] L(-M_\chi^2, m_i^2, m_i^2)+{1\over 2}\left ({R\over 1-R}-1\right )\times   \nonumber\\
&\times& \left.   \sum_{i\, j}^{N_f/2}\left [  {m_i^2 m_j^2 \over 3M_W^4}- \left ({1\over 3} - {m_i^2 +m_j^2 \over 2M_W^2} \right  ) \log \,{m_i m_j \over M_W^2} + \right.\right.\nonumber\\
&+&\left. \left.{(m_i^2 -m_j^2)^3 \over 12 M_W^6} \log \,{m_i^2 \over m_j^2} +\left ({1\over 6} -{1\over 12}{m_i^2 +m_j^2 \over M_W^2} -\right.   \right.   \right.\\
&-&\left. \left. \left.  {1\over 12}{(m_i^2 -m_j^2)^2 \over M_W^4} \right  ) {1\over M_W^2} L(-M_W^2, m_i^2, m_j^2) \right ]   K_{ij} K_{ij}^+\right \}-2\Gamma^{3\chi} (tadpoles).\nonumber
\end{eqnarray}  
It is easy to check that the contribution of the tadpole diagrams is the following one 
\begin{eqnarray} 
\lefteqn{\frac{\delta M_\chi^2}{M_\chi^2} (tadpoles) = \frac{ig^2}{16\pi^2} \left \{ \left  [ {9\over 4} r_W +9r_W^{-1}+{9\over 2R} r_Z^{-1} -{6\over M_W^2 M_\chi^2} Tr m_i^4\right  ] P - \right.  }\nonumber\\ &-&\left.   {9\over 8}r_W +{9\over 8}r_W \log r_W -{3\over 2}r_W^{-1} -{3\over 4R}r_Z^{-1} -{9\over 4R}r_Z^{-1} \log R +{3\over 2M_W^2 M_\chi^2} Tr\  m_i^4 - \right. \nonumber\\ 
&-& \left. {3\over M_W^2 M_\chi^2} Tr\ m_i^4 \log {m_i^2\over M_W^2} \right \}. \end{eqnarray} 
Using the coupling constants values from the Table I we obtain 
\begin{eqnarray}
\frac{\delta M_W^2}{M_W^2} (tadpoles) &=& {2\over 3} \frac{\delta M_\chi^2}{M_\chi^2} (tadpoles)\\
\frac{\delta M_Z^2}{M_Z^2} (tadpoles) &=& {2\over 3} \frac{\delta M_\chi^2}{M_\chi^2} (tadpoles)
\end{eqnarray} and \begin{equation}
\Gamma^{3\chi} (tadpoles) = {g\over 2M_W} \Pi^\chi (tadpoles).
\end{equation} 
Thus, one has the following additional terms in the expression for the counterterm $\Gamma^{c.t.}$ in the unitary gauge: 
\begin{eqnarray}
\Gamma^{c.t.} (tadpoles) &=& -{3g\over 2} {M_\chi^2 \over M_W}\times {2\over 3} \frac{\delta M_\chi^2}{M_\chi^2} (tadpoles)=\nonumber\\
&=& -{3g\over 2}{M_\chi^2 \over M_W} \times {2\over 3} {2M_W \over gM_\chi^2} \Gamma^{3\chi} (tadpoles) = - 2 \Gamma^{3\chi}.
\end{eqnarray} 
For the counterterm ${\cal M}$ one has 
\begin{eqnarray}
{\cal M}^{c.t.} (tadpoles) &=& -{3g^2\over 4} {M_\chi^2 \over M_W^2} \times {1\over 3} \frac{\delta M_\chi^2}{M_\chi^2} (tadpoles) =\nonumber\\
&& \hspace{-15mm}-{3g^2\over 4} {M_\chi^2\over M_W^2}\times {1\over 3} {2M_W\over g} \frac{\Gamma^{3\chi} (tadpoles)}{M_\chi^2} = - {g\over 2M_W} \Gamma^{3\chi}.
\end{eqnarray}  
As a result, the renormalized vertex is 
\begin{eqnarray}
\lefteqn{\Gamma^{ren} (p^2, q^2, (p-q)^2) =\frac{ig^3}{16\pi^2}\frac{M_\chi^2}{M_W}\left \{
\left [  {3\over 4} \frac{p^2 q^2 +p^2 (p-q)^2 +q^2 (p-q)^2}{M_W^2 M_\chi^2} -{9\over 4}r_W\right   ]P + \right.    \nonumber}\\
&+&\left.  {1\over 4} \frac{p^2 +q^2 +(p-q)^2}{M_\chi^2} \left (1+{1\over 2R}\right  ) -\right.\nonumber\\
&-&\left. {1\over 4} \frac{p^2 q^2 +p^2 (p-q)^2 +q^2 (p-q)^2}{M_W^2 M_\chi^2} \left (3+{1\over 2} \log \,R\right  ) -\right.      \nonumber\\
&-&\left.  {49\over 4} +14 R +6R^2 +{25\over 16R} +{1\over 16R^2} -{3\over 2}r_W -{27\over 4}r_W^{-1} -{27\over 8R} r_Z^{-1} +\right.      \nonumber\\
&+&\left.  {3\over 8}r_W^2 -{3\over 8}r_W r_Z +{5\over 3} (1-R) Tr \, Q_i^2 - {1\over M_W^2} Tr \, \left [  {5\over 8}m_i^2 + \left (2 \vert Q_i \vert - 4 Q_i^2 \right  ) R m_i^2\right.      \right.      \nonumber\\
&+&\left. \left.  4 Q_i^2 R^2 m_i^2\right ]   - {1\over M_W^4} Tr \, \left [ {1\over 8}m_i^4 -{1\over 8}{R\over 1-R}m_i^4 -{9\over 2}r_W^{-1} m_i^4\right ]  +\right.      \nonumber\\
&+&\left.  \left ( {3\over 8}r_W -{9\over 16R}\right ) \log \,R- {1\over 1-R} \left (-{47\over 8} +{7\over 2R} -{3\over 8R^2}-{1\over 32R^3} +\right. \right.      \nonumber\\
&+&\left.   {R\over 1-R} \left (-{1\over 48}-{1\over 3} R +{17\over 12} R^2 +R^3\right  ) {1\over M_W^2} L (-M_Z^2, M_W^2, M_W^2) +\right. \nonumber\\
&+&\left.  {R\over 1-R}\left (-{1\over 4}+{1\over 12} r_Z -{1\over 48}r_Z^2\right  ) {1\over M_W^2} L(-M_Z^2, M_Z^2, M_\chi^2) +\right.\nonumber\\
&+&\left. {1\over 1-R} \left (-{1\over 4} +{1\over 2}R +\right.      \right.      \nonumber\\
&+&\left.  \left.  {1\over 12}r_W-{1\over 6}r_Z +{1\over 24}r_W r_Z - {1\over 48} r_W^2\right  ) {1\over M_W^2} L(-M_W^2, M_W^2, M_\chi^2) +\right. \nonumber\\
&+&\left.  {1\over 1-R} \left ({25\over 12} - {11\over 6}R -2R^2 - {7\over 24R} -{1\over 48 R^2} \right  ) {1\over M_W^2} L(-M_W^2, M_W^2, M_Z^2) +\right. \nonumber\\
&+&\left.  \left (-{1\over 4}-{1\over 8}r_W^{-1} +{3\over 2}r_W^{-2} -{9\over 2} {1\over r_W^2 (r_W-4)}\right  ){1\over M_W^2} L(-M_\chi^2, M_W^2, M_W^2)+\right.      \nonumber\\
&+&\left.  \left (-{1\over 8} -{1\over 16}r_Z^{-1} + {3\over 4} r_Z^{-2} -{9\over 4}{1\over r_Z^2 (r_Z-4)}\right  ){1\over M_W^2} L(-M_\chi^2, M_Z^2, M_Z^2) +\right.         \nonumber\\
&+&\left. \left.  {9\over 16}r_Z -{3\over 16}r_Z^2 +{1\over 32}r_Z^3 \right )\log \,R +\right.\nonumber\\
&+&\left. \left (-{9\over 16} + 
{3\over 16}r_W -{3\over 16} r_Z -{1\over 32}r_W^2 +{1\over 32}r_W  r_Z +{1\over 32}r_Z^2\right  )\times \right. \nonumber\\
&\times&\left.  r_W \log \,r_W - {9\over 8M_W^2} L(-M_\chi^2, M_\chi^2, M_\chi^2)- \right.\nonumber\\
&-&\left. {1\over M_W^2} Tr \, \left [ {1\over 24} {R\over 1-R} -\left ({1\over 6}\vert Q_i \vert -{1\over 3} Q_i^2 \right  ) R-{1\over 3} Q_i^2 R^2 -\right.      \right.      \nonumber\\
&-&\left.  Tr \, \left [  {3\over 8}\left (1-{R\over 1-R}\right  ){m_i^2 \over M_W^2} +{1\over 8}{1\over 1-R}-
{1\over 2}\vert Q_i\vert -{3m_i^4\over M_W^2 M_\chi^2} \right ]   \log \,{m_i^2 \over M_W^2} +\right.      \nonumber\\
&+&\left. \left [  -{3\over 2}r_W^{-1} -\frac{q^2 +(p-q)^2-p^2}{4M_\chi^2}+ {p^2 (q^2 +(p-q)^2)\over 8M_W^2 M_\chi^2}\right ]   {1\over 2p^2}L(p^2, M_W^2, M_W^2)+\right.      \nonumber\\
&+&\left.  \left [  -{3\over 2} r_W^{-1} - \frac{p^2 +(p-q)^2 -q^2}{4M_\chi^2}+\frac{q^2 (p^2 +(p-q)^2)}{8M_W^2 M_\chi^2}\right ]  {1\over 2q^2} L(q^2, M_W^2, M_W^2)+\right.      \nonumber\\
&+&\left. \left [ -{3\over 2}r_W^{-1}-{p^2 +q^2 -(p-q)^2 \over 4M_\chi^2}+{(p-q)^2 (p^2+q^2)\over 8M_W^2 M_\chi^2}\right  ] \times \right.\nonumber\\
&\times&\left.
{1\over 2(p-q)^2} L((p-q)^2, M_W^2, M_W^2)+\right.      \nonumber\\
&+&\left.  \left [  -{3\over 4R} r_Z^{-1} - {1\over 8R} \frac{q^2 +(p-q)^2 -p^2}{M_\chi^2}+ \frac{p^2 (q^2 +(p-q)^2)}{16M_W^2 M_\chi^2}\right ]\times
\right.\nonumber\\
&\times& \left.{1\over 2p^2} L(p^2, M_Z^2, M_Z^2)+ \right.      \nonumber\\
&+&\left.  \left [  -{3\over 4R} r_Z^{-1} -{1\over 8R} \frac{p^2 +(p-q)^2 -q^2}{M_\chi^2}+
\frac{q^2 (p^2 +(p-q)^2)}{16M_W^2 M_\chi^2}\right ]  \times\right.\nonumber\\
&\times& \left. {1\over 2q^2} L(q^2, M_Z^2, M_Z^2)+\right.      \nonumber\\
&+&\left.  \left [  -{3\over 4R} r_Z^{-1} -{1\over 8R} \frac{p^2 +q^2-(p-q)^2}{M_\chi^2}+\frac{(p-q)^2 (p^2+q^2)}{16M_W^2 M_\chi^2}\right ] 
\times\right.\nonumber\\
&\times&\left. {1\over 2(p-q)^2} L((p-q)^2, M_Z^2, M_Z^2)- \right.      \nonumber\\
&-&\left.  {9\over 16} r_W\left [  {1\over 2p^2} L(p^2, M_\chi^2, M_\chi^2) +{1\over 2q^2} L(q^2, M_\chi^2, M_\chi^2) +\right. \right.\nonumber\\
&+&\left. \left. {1\over 2(p-q)^2} L((p-q)^2, M_\chi^2, M_\chi^2)\right ]   +\right.      \nonumber\\
&+&\left.  {1\over M_W^2} Tr \, \left [ {1\over 16}{R\over 1-R} -\left ({1\over 4} \vert Q_i\vert -{1\over 2}Q_i^2\right  ) R -{1\over 2}Q_i^2 R^2 -{1\over 16} {R\over 1-R}{m_i^2 \over M_Z^2} -\right.      \right.      \nonumber\\
&-&\left. \left.  {1\over 2} \left  (\vert Q_i\vert -2Q_i^2\right  ){m_i^2\over M_Z^2} R - Q_i^2 R^2 {m_i^2 \over M_Z^2}\right ]   L(-M_Z^2, m_i^2, m_i^2)+\right.      \nonumber\\
&+&\left.  {1\over M_W^2} Tr \, \left [   {3m_i^2\over 16M_\chi^2} +{21m_i^4\over 16M_\chi^4}\right ]   L(-M_\chi^2, m_i^2, m_i^2) + \right.\nonumber\\
&+&\left. {1\over M_W^2 M_\chi^2} Tr \, m_i^4 \left [  {1\over 2p^2} L(p^2, m_i^2, m_i^2) +\right. \right.      \nonumber\\
&+&\left. \left.  {1\over 2q^2} L (q^2, m_i^2, m_i^2) +{1\over 2(p-q)^2} L((p-q)^2, m_i^2, m_i^2)\right ]  -\right.      \nonumber\\
&-&\left.  {3\over 4} \left ({R\over 1-R} -1\right  ) \sum_{i\, j}^{N_f /2} \left [  {m_i m_j\over 
3M_W^2} -
\left ({1\over 3} -{m_i^2 +m_j^2 \over 2M_W^2}\right  ) \log \,{m_i m_j \over M_W^2} + \right.\right.\nonumber\\
&+&\left.\left.
\frac{(m_i^2 -m_j^2)^3}{12M_W^6} \log \,{m_i^2 \over m_j^2} + \right.     \right.      \nonumber\\
&+&\left. \left.  \left ({1\over 6}-{m_i^2 +m_j^2\over 12M_W^6} -{(m_i^2-m_j^2)^3\over 12M_W^4}\right  ){1\over M_W^2} L(-M_W^2, m_i^2, m_j^2)\right ]   K_{ij} K_{ij}^+ -\right.      \nonumber\\
&-&\left.  {27\over 8} r_W M_\chi^2 I_1 (q^2, (p-q)^2, p^2, M_\chi^2, M_\chi^2, M_\chi^2) - \right.\nonumber\\
&-&\left. \left [  6r_W^{-1}M_W^2+ \left (p^2 +q^2 + (p-q)^2\right ) r_W^{-1} +\right.      \right.      \nonumber\\
&+&\left. \left. \frac{p^4 +q^4 +(p-q)^4 -p^2 q^2 -p^2 (p-q)^2 -q^2 (p-q)^2}{2M_\chi^2} -\frac{p^2 q^2 (p-q)^2}{4M_W^2 M_\chi^2}\right  ] \times \right.      \nonumber\\
&\times&\left.   I_1 (q^2, (p-q)^2, p^2, M_W^2, M_W^2, M_W^2) - \right.\nonumber\\
&-&\left.
\left [ {3\over R} M_Z^2 r_Z^{-1} +{1\over 2R} \left (p^2+q^2 +(p-q)^2\right  ) r_Z^{-1} +\right.      \right.      \nonumber\\
&+&\left.  \left.   {1\over 4R} \frac{p^4 +q^4  +(p-q)^4 -p^2 q^2 -p^2 (p-q)^2 -q^2 (p-q)^2}{M_\chi^2} - \frac{p^2 q^2 (p-q)^2}{8M_W^2 M_\chi^2}\right  ] \times \right.       \nonumber\\
&\times&\left.   I_1 (q^2, (p-q)^2, p^2, M_Z^2, M_Z^2, M_Z^2) +\right.\nonumber\\
&+&\left. {1\over M_W^2 M_\chi^2} Tr \, m_i^4 \left (4m_i^2 +{1\over 2} (p^2 +q^2 +(p-q)^2) \right )\times \right.       \nonumber\\
&\times&\left.  I_1 (q^2, (p-q)^2, p^2, m_i^2, m_i^2, m_i^2) \right \}.
\end{eqnarray}  
\subsection{Box diagrams.}

\begin{figure}[tbp] 
  \centering
  \includegraphics[bb=59 436 374 617,width=5.67in,height=3.25in,keepaspectratio]{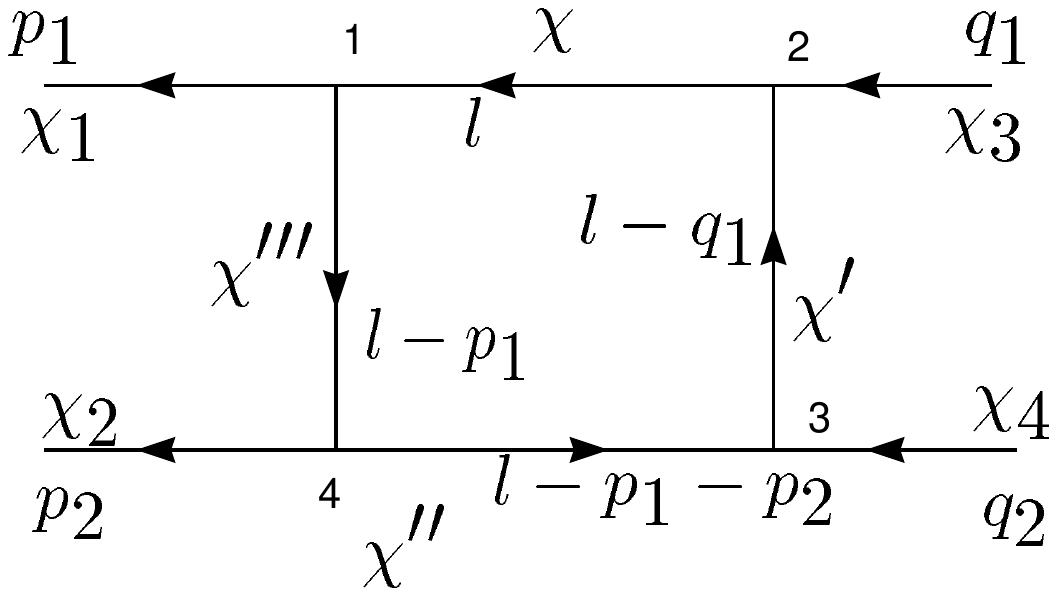}
  \caption{}
  \label{fig:FIG-H10}
\end{figure}

\begin{equation}
i{\cal M} (p_1^2, p_2^2, q_1^2, q_2^2,s ,t) = \frac{if_1^{3\chi} f_2^{3\chi} f_3^{3\chi} f_4^{3\chi}}{16\pi^2} I_2 (q_1^2, q_2^2, p_2^2, p_1^2, s, t; M_\chi^2, M_{\chi^{\prime}}^2,
M_{\chi^{\prime\prime}}^2, M_{\chi^{\prime\prime\prime}}^2),
\end{equation}

\begin{figure}[tbp] 
  \centering
  \includegraphics[bb=59 436 374 617,width=5.67in,height=3.25in,keepaspectratio]{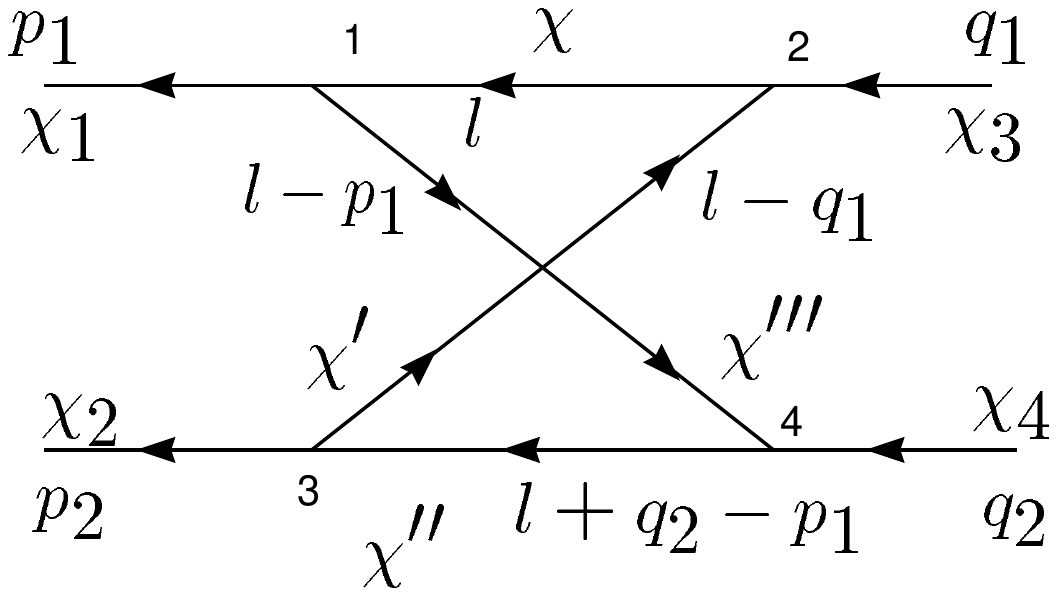}
  \caption{}
  \label{fig:FIG-H11}
\end{figure}

\begin{equation}
i{\cal M} (p_1^2, p_2^2, q_1^2, q_2^2,s ,t) = \frac{if_1^{3\chi} f_2^{3\chi} f_3^{3\chi} f_4^{3\chi}}{16\pi^2} I_2 (q_1^2, p_2^2, q_2^2, p_1^2, u, t; M_\chi^2, M_{\chi^{\prime}}^2,
M_{\chi^{\prime\prime}}^2, M_{\chi^{\prime\prime\prime}}^2).
\end{equation}

The analitical result of the diagram (Fig. 12)
\begin{figure}[tbp] 
  \centering
  \includegraphics[bb=69 449 368 617,width=5.67in,height=3.18in,keepaspectratio]{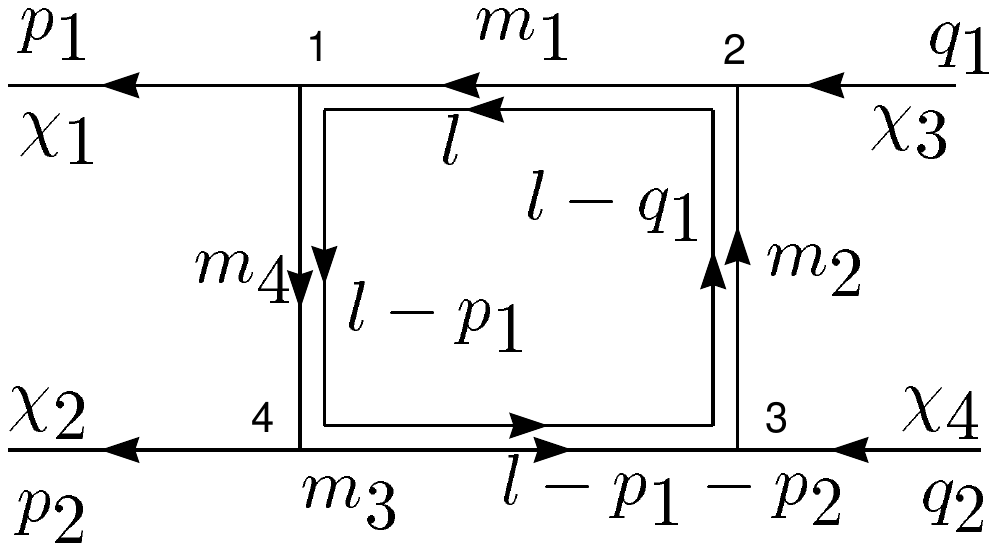}
  \caption{}
  \label{fig:FIG-H12}
\end{figure}
and the similar diagram with the opposite lepton current direction is
\begin{eqnarray}
&&i {\cal M} (p_1^2, p_2^2, q_1^2, q_2^2,s ,t) = -  \frac{if_1^{\chi} f_2^{\chi} f_3^{\chi} f_4^{\chi}}{4\pi^2}\left \{ -2B_5 P - {B_5\over 2}- {B_5\over 2}I_0 (s, m_1^2, m_3^2) -\right.      \nonumber\\
&-&\left.  {B_5\over 2} I_0 (t, m_2^2, m_4^2)+{1\over 2} \left [  B_5 \left (q_1 q_2 -m_2^2\right  ) -B_6 m_1 m_2 -B_7 m_1 m_3 - B_9 m_2 m_3 \right ] \times\right.       \nonumber\\
&\times&\left.   I_1 (q_1^2, q_2^2, s, m_1^2, m_2^2, m_3^2) + \right. \nonumber\\
&-&\left.  {1\over 2}\left [  B_5 \left (q_1 p_1 +m_1^2\right  ) -B_6 m_1 m_2 -B_8 m_1 m_4 -B_{10} m_2 m_4\right ]   I_1 (q_1^2, t, p_1^2,  m_1^2, m_2^2, m_4^2)+\right.      \nonumber\\
&+&\left. {1\over 2} \left [  B_5 \left (p_1 p_2 -m_4^2\right  ) -B_7 m_1 m_3 -B_8 m_1 m_4 -B_{11} m_3 m_4\right ]   I_1 (s, p_2^2, p_1^2, m_1^2, m_3^2, m_4^2 ) +\right.      \nonumber\\
&-&\left.  {1\over 2} \left [  B_5 \left (q_2 p_2 -m_3^2\right  ) -B_9 m_2 m_3 -B_{10} m_2 m_4 - B_{11}m_3 m_4 \right ]   I_1 (q_2^2, p_2^2, t, m_2^2, m_3^2, m_4^2) +\right.      \nonumber\\
&+&\left.  {1\over 4} \left [  B_5 \left ( (q_1^2+m_1^2 +m_2^2 ) (p_2^2 +m_3^2 +m_4^2) -(s+m_1^2 +m_3^2)(t+m_2^2 +m_4^2) + \right.      \right.      \right.    \nonumber\\
&+&\left. \left. \left.   (p_1^2 +m_1^2 +m_4^2) (q_2^2 +m_2^2 +m_3^2)  +2B_6 m_1 m_2 (p_2^2 +m_3^2 +m_4^2) +\right.\right.\right.\nonumber\\
&+&\left.\left.\left.  2B_7 m_1 m_3 (t+m_2^2 +m_4^2) +\right. \right.      \right.      \nonumber\\
&+& \left. \left. \left.   2B_8 m_1 m_4 (q_2^2 +m_2^2 +m_3^2)+2B_9 m_2 m_3 (p_1^2 +m_1^2 +m_4^2) +\right.\right.\right.\nonumber\\
&+&\left.\left.\left. 2B_{10}m_2 m_4  (s+m_1^2 +m_3^2 ) + \right.      \right.      \right.      \\
&+&\left. \left. \left.  2B_{11} m_3 m_4  (q_1^2 +m_1^2 +m_2^2\right  ) + \right.\right.\nonumber\\
&+&\left. \left. 4 B_{12} m_1 m_2 m_3 m_4 \right  ]  I_2 (q_1^2, q_2^2, p_2^2, p_1^2, s, t; m_1^2, m_2^2, m_3^2, m_4^2)\right \},
\nonumber 
\end{eqnarray}
where
\begin{eqnarray}
B_5&=&1 - b_1 b_2 + b_1 b_3  - b_1 b_4 - b_2 b_3 + b_2 b_4 - b_3 b_4+ b_1 b_2 b_3 b_4,\nonumber\\
B_6&=&1 + b_1 b_2 + b_1 b_3  - b_1 b_4 + b_2 b_3 - b_2 b_4 - b_3 b_4 - b_1 b_2 b_3 b_4,\nonumber\\
B_7&=&1 + b_1 b_2 - b_1 b_3  - b_1 b_4 - b_2 b_3 - b_2 b_4 + b_3 b_4+ b_1 b_2 b_3 b_4,\nonumber\\
B_8&=&1 + b_1 b_2 - b_1 b_3  + b_1 b_4 - b_2 b_3 + b_2 b_4 - b_3 b_4 - b_1 b_2 b_3 b_4,\nonumber\\
B_9&=&1 - b_1 b_2 - b_1 b_3  - b_1 b_4 + b_2 b_3 + b_2 b_4 + b_3 b_4 -  b_1 b_2 b_3 b_4,\nonumber\\
B_{10}&=&1 - b_1 b_2 - b_1 b_3  + b_1 b_4 + b_2 b_3 - b_2 b_4 - b_3 b_4+ b_1 b_2 b_3 b_4,\nonumber\\
B_{11}&=&1 - b_1 b_2 + b_1 b_3  + b_1 b_4 - b_2 b_3 - b_2 b_4 + b_3 b_4 - b_1 b_2 b_3 b_4,\nonumber\\
B_{12}&=&1 + b_1 b_2 + b_1 b_3  + b_1 b_4 + b_2 b_3 + b_2 b_4 + b_3 b_4+ b_1 b_2 b_3 b_4,
\end{eqnarray}

The diagram (Fig. 13)
\begin{figure}[tbp] 
  \centering
  \includegraphics[bb=70 441 376 624,width=5.67in,height=3.39in,keepaspectratio]{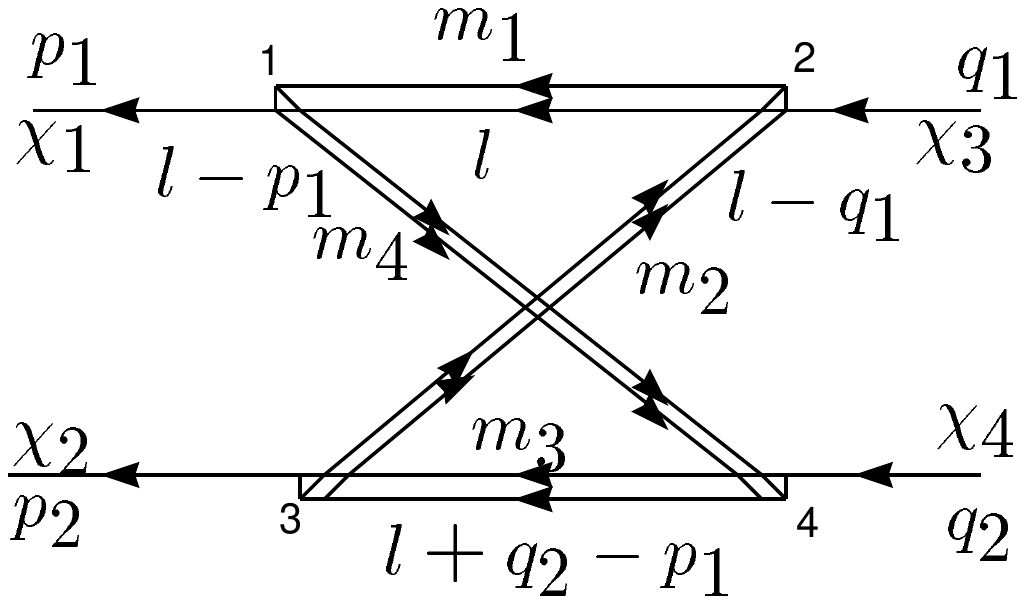}
  \caption{}
  \label{fig:FIG-H13}
\end{figure}
and the analogous diagram with the opposite lepton current direction
are described by the above expression but with the substitution $p_2\Leftrightarrow -q_2$.

The diagram (Fig. 14)
\begin{figure}[tbp] 
  \centering
  \includegraphics[bb=61 450 371 624,width=5.67in,height=3.18in,keepaspectratio]{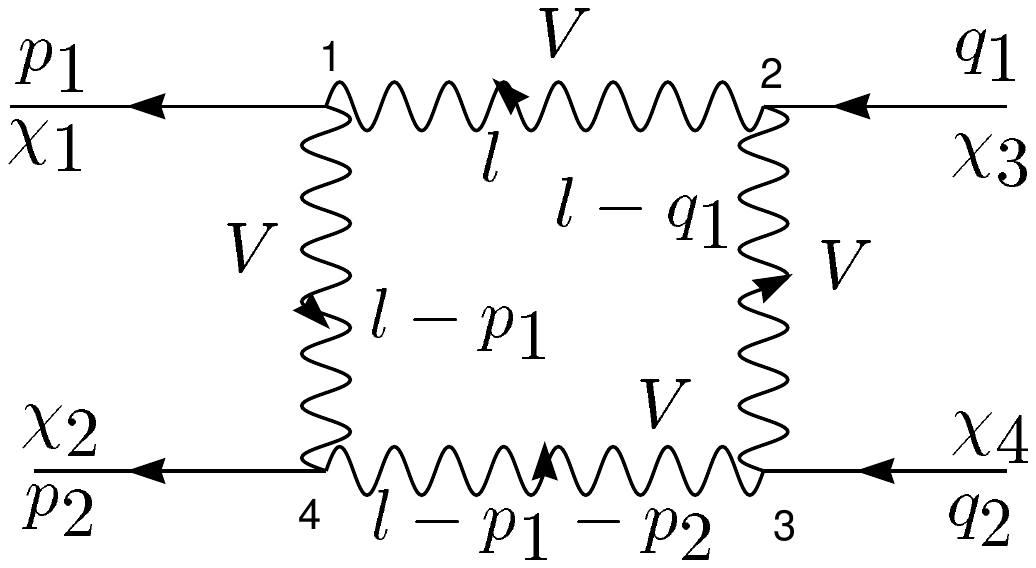}
  \caption{}
  \label{fig:FIG-H14}
\end{figure}
and the similar diagram with the opposite vector boson current direction are described by
\begin{eqnarray}
\lefteqn{ i {\cal M} (p_1^2, p_2^2, q_1^2, q_2^2, s, t) = \frac{if_1^{V\chi} f_2^{V\chi} f_3^{V\chi} f_4^{V\chi}}{16\pi^2} \left \{ \left [  {3\over 2M_V^6} \left (s+t - p_1^2 - p_2^2 - q_1^2 - q_2^2\right ) - \right.       \right.  }\nonumber\\
&-&\left.  \left.  {1\over 8M_V^8}\left ( (p_1^2 + p_2^2 + q_1^2 + q_2^2)(p_1^2 +p_2^2 +q_1^2 +q_2^2 - {s+t \over 6}) + p_1^2 q_2^2 +p_2^2 q_1^2 -\right. \right.      \right.      \nonumber\\
&-& \left.  \left. \left.  {(k_2^2 -k_1^2)(p_2^2 - p_1^2)\over 3} - {(k_1^2 - p_1^2) (k_2^2 -p_2^2)\over 3} -{s^2+t^2 -st\over 3}\right ) \right ] P+\right.      \nonumber\\
&+&\left.  \left [  3+{1\over 2M_V^2} \left (p_1^2 +p_2^2 +q_1^2 +q_2^2 \right ) +
{1\over 4M_V^4} \left (p_1^4 +p_2^4 +q_1^4 +q_2^4 +\right.      \right. \right.      \nonumber\\
&+&\left. \left. \left.  s^2 +t^2 -(s+t) (p_1^2+p_2^2+q_1^2+q_2^2) +p_1^2 q_2^2 +p_2^2 q_1^2\right ) +\right.      \right.      \nonumber\\
&+&\left. \left.  {1\over 8M_V^6} \left (p_1^2 p_2^2 (q_1^2 +q_2^2) + q_1^2 q_2^2 (p_1^2 +p_2^2) - s  (p_1^2 p_2^2 +q_1^2 q_2^2) -t (p_1^2 q_1^2 +p_2^2 q_2^2)\right ) + \right.\right.\nonumber\\
&+&\left.\left. {1\over 16M_V^8} \left  ( p_1^2 p_2^2 q_1^2 q_2^2\right )\right ] 
 I_2 (q_1^2, q_2^2, p_2^2, p_1^2, s, t, M_V^2, M_V^2, M_V^2, M_V^2) +\right.      \nonumber\\
&+&\left.  \left [  {1\over 4M_V^4} \left ( -p_1^2 - p_2^2 +t + 2s - F_1\right ) + {1\over 8M_V^6} \left  ( t (q_1^2 +q_2^2)+s (p_1^2 + p_2^2) - st - 2q_1^2 q_2^2 -\right.      \right.      \right.      \nonumber\\
&-&\left. \left.  \left.  p_1^2 q_2^2 -p_2^2 q_1^2 + sF_1 \right ) -{q_1^2 q_2^2\over 16M_V^8} \left (p_1^2 +p_2^2 -t +F_1\right )\right ]   I_1 (q_1^2, q_2^2,s, M_V^2, M_V^2, M_V^2) +\right.      \nonumber\\
&+&\left. \left [  {1\over 4M_V^4} \left  (-q_1^2 - q_2^2 +t + 2s - F_2\right ) +{1\over 8M_V^6} \left ( t (p_1^2+p_2^2)+s (q_1^2 + q_2^2) - st -2p_1^2 p_2^2 - \right. \right. \right. \nonumber\\
&-& \left. \left. \left. p_1^2 q_2^2 -p_2^2 q_1^2 + sF_2 \right ) -
{p_1^2 p_2^2\over 16M_V^8}\left ( q_1^2 + q_2^2 - t  + F_2 \right )\right ]   I_1 (p_1^2, p_2^2, s, M_V^2, M_V^2, M_V^2) + \right. \nonumber\\
&+&\left.  \left [  {1\over 4M_V^4} (-p_2^2 - q_2^2 + s + 2t - F_3) +{1\over 8M_V^6} \left (s (p_1^2 +q_1^2) +t (p_2^2 + q_2^2) - \right.\right.\right.\nonumber\\
&-&\left. \left. \left. st - 2p_1^2 q_1^2 - p_1^2 q_2^2 - p_2^2 q_1^2 + t F_3 \right ) -\right.      \right.      \nonumber\\
&-&\left. \left.  {p_1^2 q_1^2\over 16 M_V^8} (p_2^2 + q_2^2 -s +  F_3)\right ]   I_1 (q_1^2, p_1^2, t, M_V^2, M_V^2, M_V^2) +\right.\nonumber\\
&+&\left [  {1\over 4M_V^4} \left ( -p_1^2 -q_1^2+ s + 2t - F_4 \right ) +\right. \nonumber\\
&+&\left. \left. {1\over 8M_V^6} \left ( s(p_2^2 +q_2^2) +t (p_1^2+q_1^2) - st - 2p_2 q_2 -p_1^2 q_2^2 -p_2^2 q_1^2 + t F_4 \right ) -\right.\right.\nonumber\\
&-&\left. \left. {p_2^2 q_2^2\over 16 M_V^8} \left (p_1^2 + q_1^2 -s + F_4 \right ) \right ]  I_1 (q_2^2, p_2^2, t, M_V^2, M_V^2, M_V^2) +\right.\nonumber\\
&+&\left.\left [  {1\over 4M_V^4} \left (1 + F_5 + F_6 \right ) +{1\over 8M_V^6} \left ({s+t+u \over 2} -2q_1^2 + p_1 q_2 +p_2^2 - q_2^2 - p_1^2 -\right.      \right.      \right.  \nonumber\\
&-&\left. \left. \left.  s F_5 - t F_6\right ) + {q_1^2\over 16M_V^8} \left ({s+t+u \over 2} + p_2 q_1 -p_1^2 -p_2^2 - \right.\right.\right.\nonumber\\
&-&\left.\left. \left. q_1^2 - q_2^2 + q_2^2 F_5 + p_1^2 F_6\right )\right ]   I_0 (q_1^2, M_V^2, M_V^2) +\right.      \nonumber\\
&+&\left. \left  [ {1\over 4M_V^4} \left ( 1 - F_7 - F_8\right ) +{1\over 8M_V^6} \left ( {s+t+u \over 2} -2q_2^2 +p_2 q_1 + p_1^2 -
\right.\right.\right.\nonumber\\
&-&\left.\left.\left. q_1^2 - p_2 - s F_7 - t F_8 \right ) +\right.      \right.      \nonumber\\
&+&\left. \left.   {q_2^2\over 16 M_V^8} \left (  {s+t+u\over 2} + p_2 q_1 - p_1^2 -p_2^2 -q_1^2 -q_2^2 + q_1^2 F_7 + p_2^2 F_8 \right )\right  ] I_0 (q_2^2, M_V^2, M_V^2)  \right.\nonumber\\
&+&\left. 
\left [  {1\over 4M_V^4} \left (1+ F_9 +F_{10}\right ) +
{1\over 8M_V^6} \left ( {s+t+u \over 2}- 2p_1^2 + p_2 q_1 +\right.\right.\right.\nonumber\\
&+&\left.\left.\left.
 q_2^2 -p_2^2 - q_1^2 - s F_9 - t F_{10}\right ) +\right.\right.\nonumber\\
&+&\left.\left.
{p_1^2\over 16 M_V^8} \left ({s+t+u \over 2} +p_2 q_1 - p_1^2 -p_2^2 - q_1^2 - q_2^2 + q_1^2 F_9 + p_2^2 F_{10}\right ) \right ]   \times \right.       \nonumber\\
&\times&\left.  I_0 (p_1^2, M_V^2, M_V^2) +\left  [ {1\over 4M_V^4} \left (1 + F_{11} + F_{12}\right ) +\right.      \right.       \nonumber\\
&+&\left. \left.   {1\over 8M_V^6} \left ( {s+t+u\over 2} - 2p_2^2 +p_1 q_2 + q_1^2 -p_1^2 - q_2^2 - s F_{11} - t F_{12}\right ) + \right.\right.\nonumber\\
&+&\left.\left.
{p_2^2\over 16 M_V^8}\left  ( {s+t+u \over 2} +p_1 q_2 - p_1^2 -p_2^2 - \right. \right. \right. \nonumber\\
&-& \left. \left. \left. q_1^2 - q_2^2 +  p_1^2 F_{11} + q_2^2 F_{12}\right  ) \right  ] I_0 (p_2^2, M_V^2, M_V^2) + \right. \nonumber\\
&+&\left.  \left [  - {1\over 4M_V^4} \left ( 2 - F_{13} - F_{14} \right ) + {1\over 8M_V^6} \left ( {11\over 6} s + {5\over 3} t
- {5\over 6} p_1^2 - {5\over 6} p_2^2 - {5\over 6} q_1^2 - {5\over 6} q_2^2 -\right.      
\right.      \right.       \nonumber\\ 
&-&\left. \left. \left.  \frac{( k_2^2 -k_1^2 )( p_2^2 -p_1^2 ) }{6s} - s F_{13} - s F_{14} \right ) -\right.\right.\nonumber\\
&-&\left.\left.
- {1\over 16M_V^8}\left ( p_1^2 p_2^2 +q_1^2 q_2^2 +{q_1^2 p_2^2 + p_1^2 q_2^2 \over 2} - 
{s (p_1^2 + 
p_2^2 + q_1^2 + q_2^2 )\over 6} - \right.\right.\right.\nonumber\\
&-&\left.\left.\left.
\frac{(k_1^2 - p_1^2) (k_2^2 -p_2^2)}{3} - {st +2s^2 \over 6} + q_1^2 q_2^2 F_{13} + p_1^2 p_2^2 F_{14}  \right ] I_0 (s, M_V^2, M_V^2) - \right. \right.     \nonumber\\
&-&\left.   \left [  - {1\over 4M_V^4} \left ( 2 - F_{15} -  F_{16} \right )+{1\over 8M_V^6} \left ( 
{11\over 6} s + {5\over 3} t -{5\over 6} p_1^2 -{5\over 6} p_2^2 - {5\over 6} q_1^2 - {5\over 6} q_2^2 +\right.\right.\right.\nonumber\\
&+&\left.\left.\left. 
\frac{(k_1^2 -p_1^2) (k_2^2 -p_2^2)}{6t} - 
 t F_{15} - t F_{16}\right  ) - \right.       \right. \nonumber\\
&-& \left.  \left.   {1\over 16 M_V^8} \left  ( p_1^2 q_1^2 +p_2^2 q_2^2 +{p_1^2 q_2^2 +p_2^2 q_1^2 \over 2} +{t (p_1^2 +p_2^2 +q_1^2 +q_2^2) \over 6} - 
\right.\right.\right.\nonumber\\
&-&\left. \left. \left. {(k_1^2 - p_1^2) (k_2^2 -p_2^2)\over 3} -{2t^2 + st\over 6} - 
p_1^2 q_1^2 F_{15} - p_2^2 q_2^2 F_{16}\right )\right ]   \times\right.      \nonumber\\
&\times&\left.  I_0 (t, M_V^2, M_V^2) + {1\over 16 M_V^6}\log {M_V^2 \over M_W^2}\left [  s+t -{14\over 3}p_1^2 - {14\over 3} p_2^2 -{14\over 3} q_1^2 - {14\over 3} q_2^2 + \right.\right.\nonumber\\
&+&\left. \left. \frac{(k_2^2 -k_1^2) (p_2^2 -p_1^2)}{3s} + 
\frac{(k_1^2 -p_1^2)(k_2^2 -p_2^2)}{3t} \right  ]+\right.      \nonumber\\
&+&\left.  {1\over 4M_V^6} \left [  p_1^2 +  p_2^2 +  q_1^2 +  q_2^2\right ] +{1\over 16 M_V^8} \left [  {s^2 +4st +t^2\over 18} +\right.      \right.\\
&+&\left. \left.  {(k_2^2 -k_1^2)(p_2^2 -p_1^2) + (k_1^2 -p_1^2)(k_2^2 -p_2^2)\over 18} - {(s+t) (p_1^2 +p_2^2 +k_1^2 +k_2^2)\over 18}\right ]  \right \}.
\nonumber
\end{eqnarray}
The result from Fig. 15
\begin{figure}[tbp] 
  \centering
\includegraphics[scale=0.6]{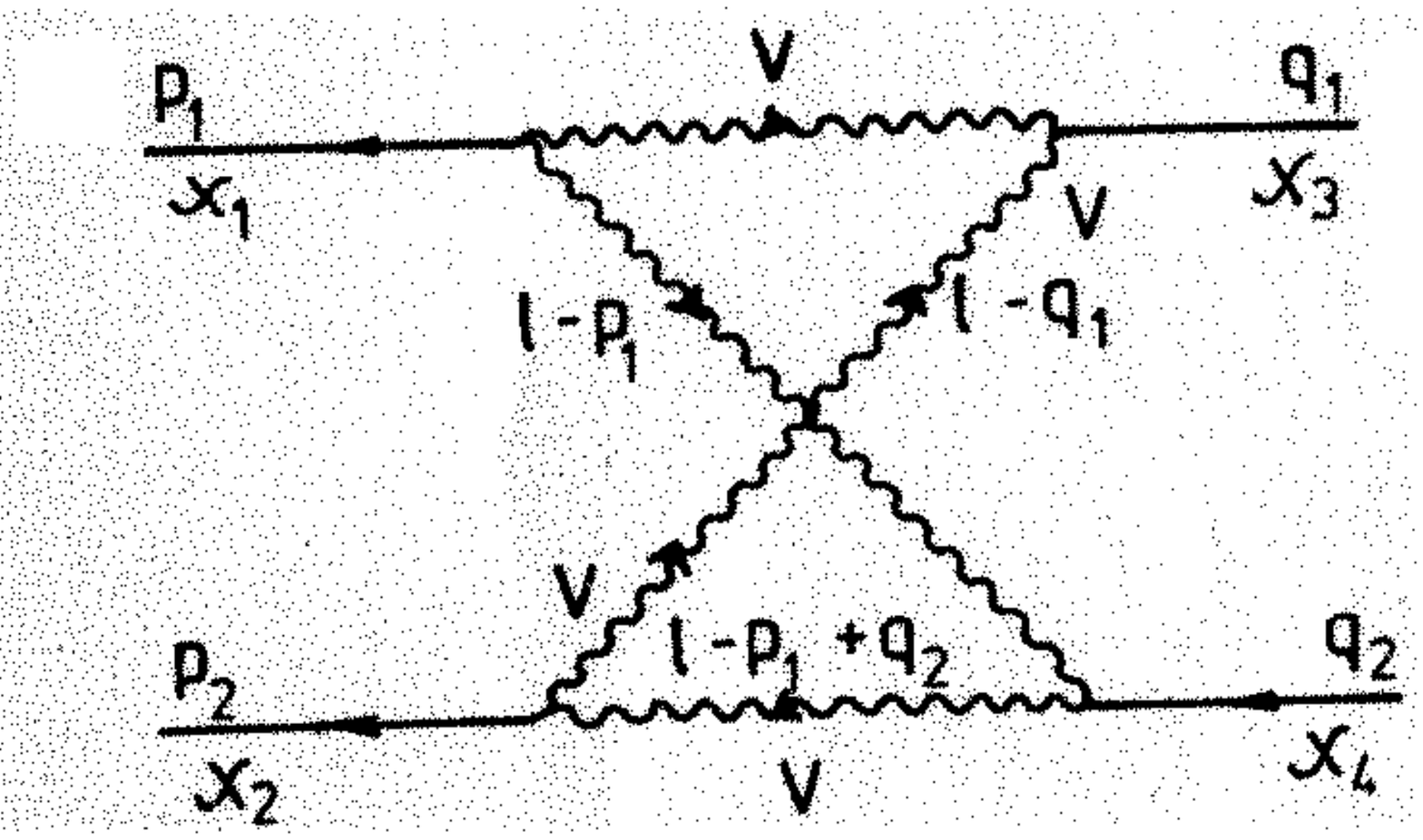} 
  \caption{}
  \label{fig:FIG-H15}
\end{figure}
and  the analogous diagram with the opposite vector boson current direction
are obtained from the above expression but with the substitution $p_2 \Leftrightarrow -q_2$.

Here,
\begin{eqnarray}
&&F_1 = \frac{1}{s^2 +q_1^4 +q_2^4 -2s q_1^2 -2s q_2^2 -2q_1^2 q_2^2}\times\\
&& \hspace{-10mm}\left \{(t+q_1^2 -p_1^2)\left [ (s+q_1^2 -q_2^2) (q_1^2 -q_2^2) - 2sq_1^2 \right ]
+  q_1^2 (p_2^2 -p_1^2 +q_1^2 -q_2^2 ) (s+q_2^2 -q_1^2)\right\}\nonumber\\
&&F_2 = \frac{1}{s^2 +p_1^4 +p_2^4 -2sp_1^2 -2sp_2^2 -2p_1^2 p_2^2}\times\\
&&\hspace{-10mm}\left\{(t+ p_1^2 - q_1^2) \left [ (s+p_1^2 -p_2^2) (p_1^2 -p_2^2) - 2sp_1^2 \right ]
+ p_1^2 (p_1^2 -p_2^2 +q_2^2 -q_1^2) (s+p_2^2 -p_1^2) \right\}\nonumber\\
&&F_3 = \frac{1}{t +p_1^4 +q_1^4 -2p_1^2 q_1^2 -2tp_1^2 -2tq_1^2}\times\\
&&\hspace{-10mm}\left\{ (s +q_1^2 - q_2^2) \left [ (t+q_1^2 -p_1^2) (q_1^2 -p_1^2) -2tq_1^2 \right ]
+ q_1^2 (p_2^2 - p_1^2  +q_1^2 - q_2^2) (t +p_1^2 - q_1^2)\right\}
\nonumber\\
&&F_4 = \frac{1}{t^2 +p_2^4 +q_2^4 -2tp_2^2 -2t q_2^2 -2p_2^2 q_2^2}\times\\ 
&&\hspace{-10mm}\left\{ (s +q_2^2 -q_1^2) \left [ (t +q_2^2 - p_2^2) (q_2^2 - p_2^2) -2tq_2^2 \right ]
+ q_2^2 (p_1^2 -p_2^2 + q_2^2 -q_1^2) (t +q_2^2 -p_2^2) \right\}
\nonumber
\end{eqnarray}
\begin{eqnarray}
F_5 &=& \frac{ (t+q_1^2 -p_1^2) (s+q_1^2 -q_2^2) + 2q_1^2 (p_1^2 - p_2^2 +q_2^2 - q_1^2)}{s^2 +q_1^4 +q_2^4 -2sq_1^2 -2sq_2^2 -2q_1^2 q_2^2}\\
F_6 &=& \frac{ (s+ q_1^2 -q_2^2) (t +q_1^2 - p_1^2) + 2q_1^2 (p_1^2 - p_2^2 +q-2^2 - q_1^2)}{t^2 + p_1^4 +q_1^4 - -2tp_1^2 -2t q_1^2 - 2p_1^2 q_1^2}\\
F_7 &=& \frac{ (t + q_1^2  - p_1^2) (s + q_2^2 - q_1^2)  + (p_1^2 -p_2^2 +q_2^2 -q_1^2 ) (s -q_1^2 -q_2^2)}{ s^2 +q_1^4 +q_2^4 -2sq_1^2 -2sq_2^2 -2q_1^2 q_2^2}\nonumber\\
&&\\
F_8 &=& \frac{ (s+ q_2^2 - q_1^2) (t+ q_2^2 - p_2^2) + 2q_2^2 (p_2^2 - p_1^2 +q_1^2 - q_2^2)}{t^2 +p_2^4 +q_2^4 - 2tp_2^2 -2tq_2^2 -2p_2^2 q_2^2}\\
F_9 &=& \frac{ (t + p_1^2 - q_1^2) (s+ p_1^2 -p_2^2) + 2p_1^2 (p_2^2 -p_1^2 +q_1^2 -q_2^2)}{s^2 +p_1^4 +p_2^4 -2sp_1^2 -2sp_2^2 -2p_1^2 p_2^2}\\
F_{10} &=& \frac{ (s+ q_1^2 - q_2^2) (t +p_1^2 - q_1^2) + (t - q_1^2 -p_1^2) (p_2^2 -p_1^2 +q_1^2 - q_2^2)}{t^2 +p_1^4 +q_1^4 - 2tp_1^2 -2t q_1^2 - 2p_1^2 q_1^2}\nonumber\\&&\\
F_{11} &=& \frac{ (t+p_1^2 -q_1^2) (s+p_2^2 -p_1^2) + (s- p_1^2 -p_2^2) ( p_2^2 - p_1^2 + q_1^2 - q_2^2)}{ s^2 + p_1^4 +p_2^4 - 2sp_1^2 -2sp_2^2 -2p_1^2 p_2^2}\nonumber\\&&\\
F_{12} &=& \frac{ (s+ q_2^2 - q_1^2) ( t+p_2^2 - q_2^2) + (t -p_2^2 - q_2^2) (p_2^2 - p_1^2 + q_1^2 - q_2^2)}{t^2 + p_2^4 + q_2^4  - 2tp_2^2 - 2t q_2^2 - 2p_2^2 q_2^2}\nonumber\\&&\\
F_{13} &=& \frac{ -2s (t + q_1^2 -p_1^2)+ (s+ q_1^2 - q_2^2) ( p_1^2 - p_2^2 + q_2^2 - q_1^2 )}{s^2 + q_1^4 +q_2^4 - 2sq_1^2 - 2sq_2^2 -2q_1^2 q_2^2}\\
F_{14} &=& \frac{  -2s ( t + p_1^2 - q_1^2) + (s +p_1^2 - p_2^2) ( p_1^2 - p_2^2 + q_2^2 - q_1^2)}{ s^2 +p_1^4 +p_2^4 - 2sp_1^2 -2sp_2^4 - 2p_1^2 p_2^2}\\
F_{15} &=& \frac{  -2t (s + q_1^2 - q_2^2) + (t + q_1^2 - p_1^2) (  p_1^2 - p_2^2 + q_2^2 - q_1^2)}{t^2 +p_1^4 +q_1^4 -2tp-1^2 -2t q_1^2 - 2q_1^2 q_2^2}\\
F_{16} &=& \frac{ -2t (s + q_2^2 -q_1^2) + (t + q_2^2 - p_2^2) ( p_1^2 - p_2^2 + q_2^2 - q_1^2)}{t^2 +p_2^4 +q_2^4 -2tp_2^2 -2t q_2^2 - 2p_2^2 q_2^2}
\end{eqnarray}
and
\begin{eqnarray}
p_1 p_2&=&{1\over 2} \left (s-p_1^2 -p_2^2 \right ), \qquad p_1 q_2={1\over 2} \left (s+t-p_2^2-q_1^2 \right ),\nonumber\\
q_1 q_2&=& {1\over 2}\left (s-q_1^2 -q_2^2\right ), \qquad p_2 q_1 ={1\over 2} \left (s+t- p_1^2 -q_2^2\right ),\nonumber\\
p_1 q_1 &=&- {1\over 2}\left ( t-p_1^2 -q_1^2\right ),\nonumber\\
p_2 q_2 &=&- {1\over 2}\left (t - p_2^2 -q_2^2\right )
\end{eqnarray}
with $s, t, u$ being the Mandelstam variables.
$I_0, I_1, I_2$ are the scalar one-loop integrals calculated in~\cite{13}.

\section{The one-loop amplitudes.}

The box diagrams of the previous Section and the diagrams drawn at {\it Fig. 16} (see also~[13b]) 
contribute to the Higgs-Higgs amplitude up to fourth order of perturbation theory to give
\begin{eqnarray}
i{\cal M}_a &=& {9ig^2 \over 4} r_W {M_\chi^2 \over t+M_\chi^2},\\
i{\cal M}_{b,c} &=& -{3g \over 2} r_W \frac{M_W \Gamma^{ren} (p^2_{1,2}, q_{1,2}^2, t)}{t+M_\chi^2},\\
i{\cal M}_{d,e} &=& {9g^2\over 4} r_W \frac{M_\chi^2 \Pi^{ren} (p_{1,2}^2)}{(p_{1,2}^2 +M_\chi^2) (t+M_\chi^2)},\\
i{\cal M}_{f,g} &=& {9g^2\over 4} r_W \frac{M_\chi^2  \Pi^{ren} (q_{1,2}^2)}{(q_{1,2}^2 +M_\chi^2) (t+M_\chi^2)},\\
i{\cal M}_{h} &=&   {9g^2\over 4} r_W \frac{M_\chi^2  \Pi^{ren} (t)}{ (t+M_\chi^2)^2},\\
i{\cal M}_{i} &=& {9g^4\over16} r_W^2 \Pi ^{(3)} (t),\\
i{\cal M}_{j} &=& {g^4\over 4} \Pi^{(5)} (t) \vert_{M_V^2=M_W^2} +{g^4\over 4} {1\over R^2} \Pi ^{(5)}(t)\vert_{M_V^2=M_Z^2},\\
i{\cal M}_{k,l} &=& -{27 \over 16} g^4 r_W^2 M_\chi^2 \Gamma^{(1)} (p_{1,2}^2, q_{1,2}, t),\\
i{\cal M}_{m,n} &=& -{g^4\over 2} M_W^2 \Gamma^{(3)} (p_{1,2}^2, q_{1,2}, t)\vert_{M_V^2 =M_W^2} -\nonumber\\
&&-{g^4\over 2}{1\over R^2} M_Z^2  \Gamma^{(3)} (p_{1,2}^2, q_{1,2}, t)\vert_{M_V^2 =M_Z^2},\\
i{\cal M}_{o} &=& -{3ig^2\over 4} r_W,\\
i{\cal M}_{p} &=& -{3g^2\over 4} r_W \frac{\Pi^{ren} (p_1^2)}{(p_1^2 +M_\chi^2)},\\
i{\cal M}_{r} &=& -{3g^2\over 4} r_W \frac{\Pi^{ren} (p_2^2)}{(p_2^2 +M_\chi^2)},\\
i{\cal M}_{s} &=& -{3g^2\over 4} r_W \frac{\Pi^{ren} (q_1^2)}{(q_1^2 +M_\chi^2)},\\
i{\cal M}_{t} &=& -{3g^2\over 4} r_W \frac{\Pi^{ren} (q_2^2)}{(q_2^2 +M_\chi^2)}.
\end{eqnarray}



The digrams describing the interaction of two Higgs particle ($s$-channel)are presnted in~\cite{12}, a solid line corresponds to the renormalized propagator and a black circle, to the renormalized vertex.

In the framework of the renormalization scheme~\cite{8} (using the unitary gauge) the appropriate counterterm for $4\chi$ interaction is written as follows:
\begin{equation}
i {\cal M}^{c.t.} = -{3g^2 M_\chi^2\over 4M_W^2}\left [  2(Z_\chi -1)+2{\delta g\over g}+{\delta M_\chi^2\over M_\chi^2}-{\delta M_W^2\over M_W^2}\right ] .
\end{equation}
After substitution of the renormalization constants (see preceding Section) we have
\begin{eqnarray}
\lefteqn{i {\cal M}^{c.t} (p_1^2, p_2^2, q_1^2, q_2^2, s, t) = - \frac{ig^4}{16\pi^2}{3r_W \over 4}
\left  \{ \left  [ - {3\over 4}r_W -9r_W^{-1} -{9\over 2R} r_Z^{-1} +\right.\right.}\nonumber\\
&+& \left.\left. {6\over M_W^2 M_\chi^2} Tr\, m_i^4 \right  ] P+
{52\over 3} -{56 \over 3} R -8R^2 -{19\over 12R} -{1\over 12R^2} +{31\over 8} r_W -\right.\nonumber\\
&-&\left. {1\over 2}r_W^2 +{1\over 2}r_W r_Z +{21\over 2} r_W^{-1} +{21\over 4 R} r_Z^{-1}+\right.\nonumber\\
&+&\left.  (-{3\over 8} r_W +{3\over 4R}+{9\over 4R} r_Z^{-1} ) \log \,R +{1\over 1-R} (-{47\over 6} + {14\over 3R}- {1\over 2R^2} -{1\over 24 R^3} +\right.      \nonumber\\
&+&\left.   {3\over 4} r_Z -
{1\over 4}r_Z^2 +{1\over 24} r_Z^3) \log \,R + (-{3\over 4} -{1\over 4}r_W +{1\over 4}r_Z +{1\over 24}r_W^2 -
\right.\nonumber\\
&-&\left.\left. {1\over 24}r_W r_Z -{1\over 24} r_Z^2) r_W \log \,r_W - 
{20\over 9} (1-R) Tr \, Q_i^2 +\right.\right.\nonumber\\
&+& \left. {1\over 2M_W^2} {R\over 1-R} (-{1\over 12}-{4\over 3}R +{17\over 3}R^2 +4R^4 ) L(-M_Z^2, M_W^2, M_W^2) +\right. \nonumber\\
&+&\left.  {1\over 2M_W^2} {R\over 1-R} (-1+{1\over 3}r_Z -{1\over 12}r_Z^2) L(-M_Z^2, M_Z^2, M_\chi^2) +\right.       \nonumber\\
&+&\left.  {1\over 2 M_W^2} {1\over 1-R} ({25\over 3} -{22\over 3}R -8R^2 -{7\over 6R} -{1\over 12 R^2}) L(-M_W^2, M_W^2, M_Z^2) +
\right. \nonumber\\
&+&\left.  {1\over 2 M_W^2} {1\over 1-R} \left ( -1 +2R +{1\over 3}r_W -{2\over 3} r_Z +\right. \right. \nonumber\\
&+& \left. \left. {1\over 6} r_W r_Z -{1\over 12} r_W^2 \right ) L(-M_W^2, M_W^2, M_\chi^2)+\right. \nonumber\\
&+&\left.  {1\over M_W^2} (-{3\over 8} +{3\over 2} r_W^{-2}- 6 {1\over r_W^2 (r_W -4)} ) L(-M_\chi^2, M_W^2, M_W^2) +\right.      \nonumber\\
&+&\left.  {1\over M_W^2} (-{3\over 16}+{3\over 4}r_Z^{-2}- 3{1\over r_Z^2 (r_Z -4)} ) L(-M_\chi^2, M_Z^2, M_Z^2) +\right.\nonumber\\
&+&\left. {21\over 16 M_W^2} L(-M_\chi^2, M_\chi^2, M_\chi^2)+\right.      \nonumber\\
&+& \left.  {1\over M_W^2} Tr \, \left [  {1\over 3} m_i^2 +({8\over 3} \vert Q _i \vert -{16\over 3} Q_i^2) Rm_i^2 +{16\over 3} Q_i^2 R^2 m_i^2 \right ]  +\right.\nonumber\\ 
&+&\left.
{1\over M_W^4} Tr \, \left [  {1\over 6}m_i^4 -{1\over 6} {R\over 1-R} m_i^4 - {15\over 2} r_W^{-1} m_i^4\right ]   +    \right.      \nonumber\\
&-&\left.  {1\over M_W^2} Tr \, \left [  {1\over 6} {1\over 1-R} M_W^2 -{2\over 3} \vert Q_i \vert M_W^2 +{1\over 2} (1-{R\over 1-R}) m_i^2 +{3\over M_\chi^2} m_i^4\right ]   \times \right.      \nonumber\\
&\times&\left.  \log \,{m_i^2 \over M_W^2}+ {1\over M_W^2} Tr \, \left [ -{1\over 12}{R\over 1-R} +({1\over 3} \vert Q_i^2 \vert -{2\over 3} Q_i^2 ) R +{2\over 3}Q_i^2 R^2 +\right.\right.\nonumber\\
&+&\left.\left.
{1\over 12} {R\over 1-R} {m_i^2 \over M_Z^2} + ({2\over 3}\vert Q_i \vert -{4\over 3}Q_i^2) R {m_i^2 \over M_Z^2} +{4\over 3}Q_i^2 R^2 {m_i^2 \over M_Z^2}\right ]   L(-M_Z^2, m_i^2, m_i^2) -\right.      \nonumber\\
&-&\left.  {1\over M_W^2} Tr \, \left [  {m_i^2 \over 4M_\chi^2} +2{m_i^4 \over M_\chi^4} \right ]   L(-M_\chi^2, m_i^2, m_i^2)+ ({R\over 1-R} -1)\times \right.\nonumber\\
&\times& \left. \sum_{i\, j}^{N_f /2} \left [  {1\over 3} {m_i^2 m_j^2 \over M_W^4} -{1\over 3} - 
 \frac{m_i^2 + m_j^2}{2M_W^2} ) \log \,{m_i m_j \over M_W^2} +{1\over 12}{(m_i^2 -m_j^2)^3 \over M_W^6} \log \,{m_i^2 \over m_j^2}+\right.    \right. \nonumber\\
&+&\left.  \left.   ({1\over 6M_W^2} -{m_i^2 +m_j^2 \over 12 M_W^4} -{(m_i^2 -m_j^2)^2 \over 12 M_W^6} ) L(-M_W^2, m_i^2, m_j^2)\right ]   K_{ij} K^+_{ij}\right \}- \nonumber\\
&-&{g\over 2M_W} \Gamma^{3\chi} (tadpoles)\, .
\end{eqnarray}

I would also lke to note that it is not useful to use the Stuart's computer algebra programm~\cite{Stuart} since the preparation of the input data for the two-Higgs processes would consume more time than the calculations by hand.

\section{Summary.}

We calculated the Higgs-Higgs amplitude in the fourth order of perturbation theory (that is, with taking into account the one-loop corrections) in the Standard Model.  The results do not coincide in full with the Durand {\it et al.} calculations
in the Feynman gauge and within the different renormalization scheme~\cite{24}.\footnote{Recently, several authors proposed models which depended on the 4-potential, and, hence, on the gauge.}

There exists a rather large number of different variants of the SM extensions. The most of these models are characterized by enlarging of the Higgs sector, namely by introduction of two and even more Higgs boson doublets~\cite{15}-\cite{17}. The supersymmetric extension of the SM~\cite{18,19} also requires introduction of at least two doublets of the scalar particles. The interest in these models has grown up in connection with the problem of suppression of the $CP$ violation in strong interactions~\cite{20} and the problem of $B^0\bar B^0$ mixing~\cite{21} because they give the alternative theoretical solution to the problem of electroweak violation of the $CP$ invariance~\cite{17,22}. Therefore, 
in the approaching paper we are going to consider the Higgs-Higgs interaction problem in the framework of the extended variants of the SM. This is easily possible due to introduction of the parameters $a_i$, $b_i$ for the fermionic interactions.

{\bf Acknowledgements}. The author express his sincere gratitude to D. Yu. Bardin, R.~N. Faustov,  and  Yu. N. Tyukhtyaev for valuable discussions. I appreciate  very much the assistance of  V. I. Kikot' in calculations, E. Saucedo for technical help, and I thank N. B. Skachkov for putting forward the problem.

\section*{Appendix.}

The integrals $I_0$, $I_1$ and $I_2$ are taken from the t'Hooft-Veltman paper~\cite{13}
\begin{eqnarray}
\lefteqn{I_0 (q^2, M_1^2, M_2^2)=\int_{0}^{1} dx \log \,\frac{q^2 x(1-x)+M_1^2x +M_2^2 (1-x)}{M_W^2}=}\nonumber\\
&=&-2+\log{M_1 M_2\over M_W^2}-{1\over 2}{M_1^2-M_2^2 \over q^2} \log{M_1^2\over M_2^2} +{1\over 2q^2} L(q^2, M_1^2, M_2^2)
\end{eqnarray}
where
\begin{eqnarray}
&&L(q^2, M_1^2, M_2^2)=\left [ (q^2+M_1^2+M_2^2)-\right.\nonumber\\
&-&\left.4M_1^2 M_2^2\right ] \int_{0}^{1}\frac{dx}{q^2 x(1-x)+M_1^2 x +M_2^2 (1-x)},
\end{eqnarray}
see Ref.~[9a,p.440].

Next,
\begin{eqnarray}
I_1(q^2, (p-q)^2, p^2, M_1^2, M_2^2, M_3^2)&=&\int^{1}_{0} dx\int_{0}^{x} dy \left [ ax^2+by^2+cxy+\right.\nonumber\\
&+&\left.dx+ey+f\right ]  ^{-1},
\end{eqnarray}
with\\
\begin{tabular}{ll}
$a=-(p-q)^2$,&$d=M_2^2-M_3^2+(p-q)^2$\\
$b=-q^2$,&$e=M_1^2-M_2^2+2pq-q^2$\\
$c=-2(pq-q^2)$,&$f=M_3^2-i\epsilon$.
\end{tabular}

\bigskip

And,
\begin{eqnarray}
\lefteqn{I_2(q_1^2,q_2^2,p_2^2,p_1^2,s,t,M_1^2, M_2^2,M_3^2,M_4^2)=}\\
&=&\int_0^1 dx \int_0^x dy \int_0^y dz \left [ ax^2+by^2+gz^2+cxy+hxz+jyz+ \right.\nonumber\\
&+&\left.dx+ey+kz+f\right ] ^{-2},\nonumber
\end{eqnarray}
with\\
\begin{tabular}{ll}
$a=-p_2^2$,&$f=M_4^2-i\epsilon$\\
$b=-q_2^2$,&$g=-q_1^2$\\
$c=2p_2 q_2$,&$h=2p_2 q_1$\\
$d=M_3^2-M_4^2+p_2^2$,&$j=-2q_1 q_2$\\
$e=M_2^2-M_3^2+q^2-2p_2 q_2$,&$k=M_1^2-M_2^2+q_1^2+2q_1 q_2 -2q_1 p_2$.
\end{tabular}\\

The traces of  $\gamma$- matrices in an $d$-dimensional space are
\begin{eqnarray}
Tr \, \gamma_\alpha=Tr \, \gamma_5 = Tr \, \gamma_\alpha \gamma_\beta  \gamma_5 =0,\\
Tr \, \gamma_\alpha \gamma_\beta = f(d) \delta_{\alpha\beta},\\
Tr \, \gamma_\alpha \gamma_\beta \gamma_\rho \gamma_\sigma=f(d) d_{\alpha\beta\rho\sigma},\\
Tr \, \gamma_\alpha \gamma_\beta \gamma_\rho \gamma_\sigma \gamma_5=f(d) \epsilon_{\alpha\beta\rho\sigma} ,
\end{eqnarray}
where
\begin{equation} d_{\alpha\beta\rho\sigma}=\delta_{\alpha\beta}\delta_{\rho\sigma}-\delta_{\alpha\rho} \delta_{\beta\sigma}+\delta_{\alpha\sigma}\delta_{\beta\rho}.
\end{equation}
In the present work we use $f(d)=2\omega=4-2\epsilon$, $d$ is the dimension of the space in the dimensional regularization.



\begin{thebibliography}{99}

\footnotesize{

\bibitem{prog} LHC Reports, CERN-LHC-2001-001 to 007, CERN-LHC-2002-001 to 024.\\[-7mm]

\bibitem{2} R. N. Cahn and M. Suzuki, \pl {134B}{1-2}, 115-119 (1984).\\[-7mm]

\bibitem{3} M.An\-sel\-mino {\it et al.},  \pl {147B}{1-3}, 207-211 (1984).\\[-7mm]

\bibitem{23} J.A.Grifols,  \pl {264B}{1-2}, 149-153  (1991); J. A. Grifols and J. L. Diaz Cruz, UAB-FT-280-92, Dec. 1991.\\[-7mm]

\bibitem{4} G. Passarino,  \pl {156B}{3-4}, 231-235 (1985); \ibid {161B} {4-6}, 341-346  (1985).\\[-7mm]

\bibitem{5} M. Veltman, \pl {139B}{4}, 307-309 (1984).\\[-7mm]

\bibitem{6} R. Casalbuoni {\it et al.}, \ijmp {A4}{5}, 1065-1110 (1989).\\[-7mm]

\bibitem{7} M. Veltman and F. Yndurain, \np {B325}{1}, 1-17 (1989).\\[-7mm]

\bibitem{8} D. Yu. Bardin, P. Ch. Christova and O. M. Fedorenko, \np {B175}{3}, 435-461  (1980); \ibid 
{B197}{1}, 1-44 (1982).\\[-7mm]

\bibitem{9} A. Sirlin, \prd {22}{4}, 971-981  (1980).\\[-7mm]

\bibitem{10} {\it Topical Conference on Radiative Corrections in $SU(2)\times U(1)$.} Miramare-Trieste, Italy, June 6-8, 1983.\\[-7mm]

\bibitem{11} D. Yu. Bardin, {\it Precision Verifications of the Standard Theory. JINR Lectures for
Young Scientists. No. 46} P2-88-189, Dubna:JINR, 1988, pp. 22-23 [in Russian].\\[-7mm]

\bibitem{12} V. V. Dvoeglazov, V. I. Kikot' and N. B. Skachkov JINR Communications E2-90-569, E2-90-570, Dubna:JINR, 1990;  V.V.Dvoeglazov  and N.B.Skach\-kov, JINR Communications E2-91-114,E2-91-115,E2-91-179, Dubna: JINR, 1991.\\[-7mm]

\bibitem{13} G. t' Hooft and M. Veltman, \np {B153}{3-4}, 365-401 (1979).\\[-7mm]

\bibitem{14} G. Passarino and  M. Veltman, \np {B160}{1}, 151-207 (1979).\\[-7mm]

\bibitem{24} P.N. Maher,  L.Durand and  K.Riesselmann, \prd{48}{3}, 1061-1083 (1993); \ib,
1084-1096; L.  Durand, J.  M.  Johnson and J. L.  Lopez,  \prd {45}{9}, 3112-3127 (1992); L. Durand, J. M. Johnson and P. N. Maher,  
\prd {44}{1}, 127-138 (1991); L.  Durand, J.  M. Johnson and J.  L. Lopez, \prl {64}{11}, 1215-1218 (1990).\\[-7mm]

\bibitem{Marciano} W. Marciano, G. Valencia and S. Willenbrock, \prd{40}{5}, 1725R (1989); G. E. Vagenakis, Europhysics Lett. {\bf 12}, 89 (1990).\\[-7mm]

\bibitem{Lee} B. W. Lee, C. Quigg and M. B. Thacker, \prl{38}{16}, 883-885 (1977); \prd{16}{5}, 1519-1531 (1977).\\[-7mm]

\bibitem{Cornwall} J. Cornwall, D. Levin and G. Ticktopoulos, \prd{10}{4}, 1145-1167 (1974).\\[-7mm]

\bibitem{quasipoten} A. A. Logunov and A. N. Tavkhelidze, Nuovo Cimento {\bf 29}, 380-399 (1963);
V. G. Kadyshevsky, Nucl. Phys. B{\bf 6}{2}, 125-148 (1968).\\[-7mm]

\bibitem{Stuart} R. Stuart, Comp. Phys. Comm. {\bf 48}, 367 (1988); R. Stuart and A. Gognora, ibid. {\bf 56}, 337 (1990).\\[-7mm]

\bibitem{Hooft} G. t'Hooft, Nucl. Phys. B{\bf 33}{1}, 173-199 (1971); \ibid {\bf B35}{1}, 167-188 (1971).\\[-7mm]

\bibitem{15} H. Georgi, \hj {1}{1}, 155-168 (1978).\\[-7mm]

\bibitem{16} J. F. Donoghue and Ling-Fong Li, \prd {19}{3}, 945-955 (1979).\\[-7mm]

\bibitem{17} G. C. Branco, A. J. Buras and J.-M. Gerard, \np {B259}{2-3}, 306-330 (1985).\\[-7mm]

\bibitem{18} H. E. Haber and G. L. Kane, \prp {117}{2-4}, 75-263 (1985).\\[-7mm]

\bibitem{19} J. F. Gunion and H. E. Haber, \np {B272}{1}, 1-76 (1986); \ibid {B278}{3}, 449-492 (1986); \ibid { B307}{3}, 445-475  (1988).\\[-7mm]

\bibitem{20} R. D. Peccei and H. R. Quinn, \prl {38}{25}, 1440-1443  (1977); \prd {16}{6}, 1791-1797 (1977).\\[-7mm]

\bibitem{21} S. L. Glashow and E. E. Jenkins, \pl {196B}{2}, 233-236 (1987); S. Nandi, \pl {202B}{3}, 385-387  (1988).\\[-7mm]

\bibitem{22} T. D. Lee, \prd {8}{4}, 1226-1239 (1973); S. Weinberg, \prl {37}{11}, 655-661  (1976).\\[-7mm]

}

\end{thebibliography}
\end{document}